\documentclass[twocolumn]{emulateapj}
\usepackage{graphicx}
\usepackage{dcolumn}
\begin{document}

\title{ {\it Spitzer} Uncovers Active Galactic Nuclei Missed by Optical Surveys in 7 Late-type Galaxies}

\author{S. Satyapal\altaffilmark{1}, D. Vega\altaffilmark{1}, R. P. Dudik\altaffilmark{1,2}, N. P. Abel\altaffilmark{3}, \& T. Heckman\altaffilmark{4}}

\altaffiltext{1}{George Mason University, Department of Physics \& Astronomy, MS 3F3, 4400 University Drive, Fairfax, VA 22030; satyapal@physics.gmu.edu}

\altaffiltext{2}{Observational Cosmology Laboratory, NASA Goddard Space Flight Center, Greenbelt, MD 20711}

\altaffiltext{3}{Department of Physics, University of Cincinnati, Cincinnati, OH 45221}

\altaffiltext{4}{Center for Astrophysical Sciences, Department of Physics and Astronomy, The Johns Hopkins University, Baltimore, MD 21218}

\begin{abstract}

We conducted a high resolution mid-infrared spectroscopic investigation using {\it Spitzer} of 32 late-type (Sbc or later) galaxies that show no definitive signatures of Active Galactic Nuclei (AGN) in their optical spectra in order to search for low luminosity and/or embedded AGN.  These observations reveal the presence of the high ionization [NeV] 14$\mu$m and/or 24$\mu$m line in 7 sources, providing strong evidence for AGNs in these galaxies.  Taking into account the variable sensitivity of our observations, we find that the AGN detection rate based on mid-infrared diagnostics in optically normal late-type galaxies is $\sim$ 30\%, implying an AGN detection rate in late-type galaxies that is possibly 4 times larger than what optical spectroscopic observations alone suggest. We demonstrate using photoionization models with both an input AGN and an extreme EUV-bright starburst ionizing radiation field that the observed mid-infrared line ratios in our 7 AGN candidates cannot be replicated unless an AGN contribution, in some cases as little as 10\% of the total galaxy luminosity, is included. These models show that when the fraction of the total luminosity due to the AGN is low, optical diagnostics are insensitive to the presence of the AGN.  In this regime of parameter space, the mid-infrared diagnostics offer a powerful tool for uncovering AGN missed by optical spectroscopy.   The AGN bolometric luminosities in our sample inferred using our [NeV] line luminosities range from $\sim$ 3$\times$10$^{41}$ ergs s$^{-1}$ to $\sim$ 2$\times$10$^{43}$ ergs s$^{-1}$.  Assuming that the AGN is radiating at the Eddington limit, this range corresponds to a lower mass limit for the black hole that ranges from $\sim$ 3$\times$10$^3$M$_{\odot}$ to as high as $\sim$ 1.5$\times$10$^5$M$_{\odot}$.  These lower mass limits however do not put a strain on the well-known relationship between the black hole mass and the host galaxy's stellar velocity dispersion established in predominantly early-type galaxies.  Our findings add to the growing evidence that black holes do form and grow in low-bulge environments and that they are significantly more common than optical studies indicate.  

\end{abstract}


\section{Introduction}

The vast majority of {\it currently} known active galactic nuclei (AGN) in the local Universe reside in host galaxies with prominent bulges (e.g. Heckman 1980a; Keel 1983b; Terlevich, Melnick, \& Moles 1987; Ho, Filippenko, \&  Sargent 1997; Kauffmann et al. 2003). This result, together with the finding that the black hole mass, M$_{\rm BH}$, and the host galaxy's stellar velocity dispersion, $\sigma$, is strongly correlated (Gebhardt et al. 2000; Ferrarese \&  Merritt 2000), has led to the general consensus that  black hole formation and growth is fundamentally connected to the build-up of galaxy bulges.  Indeed, it has been proposed that feedback from the AGN regulates the surrounding star formation in the host galaxy (e.g. Silk \& Rees 1998; Kauffmann \& Haehnelt 2000).  However, an important outstanding question remains unresolved: {\it is a bulge in general a necessary ingredient for a black hole to form and grow}?  On the one hand, M33, the best studied nearby bulgeless galaxy, shows no evidence of a supermassive black hole (SBH), and the upper limit on the mass is significantly below that predicted by the M$_{\rm BH}$-$\sigma$ relation established in early-type galaxies (Gebhardt et al. 2001; Merritt et al. 2001).  On the other hand, both NGC 4395 (Filippenko \& Ho 2003) and POX52 (Barth et al. 2004) show no evidence for a bulge and yet do contain AGN.  However, these two galaxies have remained isolated cases of bulgeless galaxies with accreting black holes, suggesting that they are anomalies.  Indeed in the extensive Palomar optical spectroscopic survey of 486 nearby galaxies (Ho, Filippenko, \& Sargent1997; henceforth H97), there are only 9 optically identified Seyferts with Hubble type of Sc or later.  Only one galaxy of Hubble type of Scd or later in the entire survey is classified as a Seyfert (NGC 4395).  Greene \& Ho (2004, 2007) recently conducted an extensive search for broad-line AGN with intermediate mass black holes in the Fourth Data Release of the Sloan Digital Sky Survey (2004, 2007). Of the 8435 broad line AGN, they found only 174 (2\%) such intermediate mass objects, indicating that they are extremely rare.  Forty percent of these seem to be in late-type galaxies with colors consistent with morphological type Sab, although the Sloan images are of insufficient spatial resolution to extract precise morphological information.  Optical observations thus clearly suggest that AGN in late-type galaxies are uncommon.

However, determining if AGN reside in low bulge galaxies cannot be definitively answered using optical observations alone. The problem arises because a putative AGN in a galaxy with a minimal bulge is likely to be both energetically weak and deeply embedded in the center of a dusty late-type spiral.  As a result, optical emission lines will be dominated by the emission from star formation regions, severely limiting the diagnostic power of optical surveys in determining the incidence of accreting black holes in low-bulge systems. In our previous work, we demonstrated the power of mid-IR spectroscopy in detecting the low-power AGN in the Sd galaxy NGC 3621 (Satyapal et al. 2007; henceforth S07).  AGN show prominent high ionization fine structure line emission at mid-infrared wavelengths but starburst and normal galaxies are characterized by a lower ionization spectra characteristic of HII regions ionized by young stars (e.g. Genzel et al. 1998; Sturm et al. 2002; Satyapal et al. 2004).  In particular, the [NeV] 14 $\mu$m  (ionization potential 97 eV) line is not generally produced in HII regions surrounding young stars, the dominant energy source in starburst galaxies, since even hot massive stars emit very few photons with energy sufficient for the production of Ne4+.  The detection of this line in any galaxy provides strong evidence for an AGN.  

The goal of this paper is to answer the question: are AGN in late-type galaxies more common than previously thought?  If so, this would revise our understanding of the environments in which supermassive black holes (SMBHs) can form and grow and perhaps shed light on the nature of the M$_{\rm BH}$-$\sigma$ relation in low-bulge systems.  Toward this end, we conducted an archival investigation of 32 late-type galaxies observed by the {\it Spitzer} high resolution spectrograph that show no definitive optical signatures of AGN to search for low-power and/or deeply embedded AGN.

This paper is structured as follows.  In Section 2, we summarize the properties of the {\it Spitzer} archival sample presented in this paper.  In Section 3, we summarize the observational details and data analysis procedure, followed by a description of our results in Section 4.  In Section 5, we discuss the origin of the [NeV] emission , with each galaxy discussed individually, followed by a discussion of the implications of our discoveries in Section 6.  A summary of our major conclusions is given in Section 7.

\section{The Sample}

The target sources were selected from the Palomar survey of nearby bright galaxies (H97).  Of the 486 galaxies in the Palomar survey, 169 are of Hubble type of Sbc or later and classified based on their optical line ratios as ``HII'' stellar-powered galaxies or ``T2'' transition galaxies.  T2 galaxies have optical line ratios intermediate between HII galaxies and low-ionization nuclear emission-line regions (LINERs) and have no broad permitted lines (e.g. H$\alpha$) in their optical spectrum.  There is therefore no firm optical spectroscopic evidence for AGN in these galaxies.  Of these 169 galaxies, 32 were observed by the high resolution modules of the Infrared Spectrograph (IRS; Houck et al. 2004) onboard {\it Spitzer}.

Table 1 summarizes the basic properties of the galaxies in our sample.  All targets are nearby, ranging in distance from 1.6 to 43.6 Mpc.  The majority of galaxies are classified as HII galaxies; only 4 out of 32 are T2 objects.  Figure 1 shows the distribution of Hubble types for the sample. Note that there are only 2 Sd and 1 Sdm galaxy in the sample and that the majority of galaxies are of Hubble type Sc and Sbc.

%**********************************************************************
\begin{figure}[]
\noindent{\includegraphics[width=9cm]{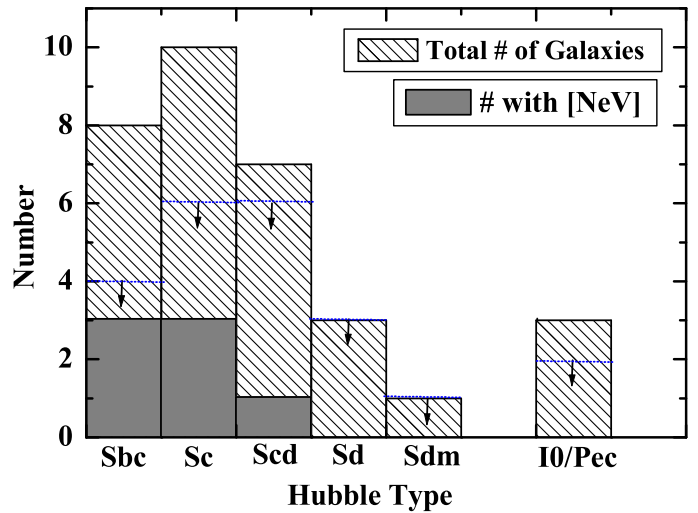}}
\caption[]{The distribution of Hubble types for the sample.  The galaxies with [NeV] detections are indicated by the filled histogram (see Section 4 for details).  Since the sensitivity of the observations varied across the sample, we also indicate with a downward arrow in Figure 1 the number of galaxies with [NeV] 14$\mu$m line sensitivity of 10$^{38}$ ergs s$^{-1}$or better.  Although the sample size is too small to make statistically meaningful statements, we note that none of the galaxies of Hubble type later than Scd in this sample display a [NeV] line.
}
\end{figure}

%**********************************************************************

\begin{table*}
\fontsize{9pt}{9pt}\selectfont
\begin{center}
\begin{tabular}{lclccccccc}
\multicolumn{10}{l}{{\bf Table 1: Properties of the Sample}}\\ 
\hline
\multicolumn{10}{l}{}\\
\multicolumn{1}{c}{Galaxy} & Distance & \multicolumn{1}{c}{Hubble} & T & M$_{B_T}^0$ & [OIII]/H$_{\beta}$ & [OI]/H$_\alpha$ & [NII]/H$_\alpha$ & [SII]/H$_\alpha$ & Optical\\

\multicolumn{1}{c}{Name} & (Mpc) & \multicolumn{1}{c}{Type} & & & & & & & Class\\

\multicolumn{1}{c}{(1)} & (2) & \multicolumn{1}{c}{(3)} & (4) & (5) & (6) & (7) & (8) & (9) & (10)\\ 
\multicolumn{9}{l}{}\\
\hline
\multicolumn{9}{l}{}\\
NGC925 & 9.4 & SAB(s)d & 7 & -19.90 & 0.84 & 0.01 & 0.22 & 0.33 & H\\
NGC1569 & 1.6 & IBm & 10 & -16.60 & 5.48 & 0.00 & 0.04 & 0.03 & H\\
NGC2276 & 36.8 & SAB(rs)c & 5  & -21.08 & 0.15 & 0.01 & 0.35 & 0.18 & H\\
NGC2903 & 6.3 & SAB(rs)bc  & 4 & -19.89 & 0.10 & 0.01 & 0.34 & 0.19 & H\\
NGC2976 & 2.1 & SAc pec  & 5 & -16.31 & 1.42 & 0.01 & 0.31 & 0.18 & H\\
NGC3034 & 5.2 & IAO spin & 90 & -19.72 & 0.36 & 0.01 & 0.56 & 0.18 & H\\
NGC3077 & 2.1 & IAO pec & 90 & -16.29 & 0.75 & 0.00 & 0.21 & 0.15 & H\\
NGC3184 & 8.7 & SAB(rs)cd  & 6 & -19.36 & 0.13 & 0.01 & 0.33 & 0.20 & H\\
NGC3198 & 10.8 & SB(rs)c  & 5 & -19.96 & 0.23 & 0.03 & 0.42 & 0.32 & H\\
NGC3310 & 18.7 & SAB(r)bc pec & 4 & -20.41 & 0.95 & 0.04 & 0.66 & 0.26 & H\\
NGC3367 & 43.6 & SB(rs)c  & 5 & -21.28 & 0.50 & 0.03 & 0.83 & 0.28 & H\\
NGC3556 & 14.1 & SB(s)cd spin  & 6 & -20.92 & 0.26 & 0.01 & 0.32 & 0.29 & H\\
NGC3726 & 17.0 & SAB(r)c  & 5 & -20.48 & 0.14 & 0.01 & 0.31 & 0.22 & H\\
NGC3938 & 17.0 & SA(s)c  & 5 & -20.32 & 1.76 & 0.14 & 0.53 & 1.67 & H\\
NGC3949 & 17.0 & SA(s)bc & 4 & -20.01 & 0.22 & 0.05 & 0.40 & 0.44 & H\\
NGC4088 & 17.0 & SAB(rs)bc  & 4 & -20.63 & 0.21 & 0.01 & 0.32 & 0.18 & H\\
NGC4192 & 17.0 & SA(s)cd spin  & 6 & -19.52 & 1.87 & 0.14 & 1.41 & 1.02 & T2\\
NGC4214 & 3.5 & IAB(s)m & H & -17.58 & 3.67 & 0.01 & 0.07 & 0.13 & H\\
NGC4236 & 2.2 & SB(s)dm  & 8 & -17.18 & 2.04 & 0.03 & 0.17 & 0.44 & H\\
NGC4254 & 16.8 & SA(s)c  & 5 & -21.03 & 0.90 & 0.02 & 0.48 & 0.23 & H\\
NGC4273 & 35.1 & SB(s)c  & 5 & -20.68 & 0.18 & 0.01 & 0.38 & 0.23 & H\\
NGC4321 & 16.8 & SAB(s)bc  & 4 & -21.15 & 0.79 & 0.11 & 1.18 & 0.48 & T2\\
NGC4414 & 9.7 & SA(rs)c & 5 & -19.31 & 0.58 & 0.14 & 0.59 & 0.50 & T2\\
NGC4490 & 7.8 & SB(s)d pec  & 7 & -19.65 & 2.55 & 0.12 & 0.25 & 0.71 & H\\
NGC4536 & 13.3 & SAB(rs)bc  & 4 & -20.04 & 0.33 & 0.03 & 0.47 & 0.36 & H\\
NGC4559 & 9.7 & SAB(rs)cd  & 6 & -20.17 & 0.35 & 0.03 & 0.42 & 0.40 & H\\
NGC4567 & 16.8 & SA(rs)bc  & 4 & -19.34 & 0.12 & 0.02 & 0.28 & 0.16 & H\\
NGC4631 & 6.9 & SB(s)d spin  & 7 & -20.58 & 1.53 & 0.03 & 0.24 & 0.23 & H\\
NGC5055 & 7.2 & SA(rs)bc  & 4 & -20.26 & 1.85 & 0.17 & 1.48 & 0.74 & T2\\
NGC5474 & 6.0 & SA(s)cd pec  & 6 & -17.59 & 1.76 & 0.02 & 0.14 & 0.27 & H\\
NGC5907 & 14.9 & SA(s)c spin  & 5 & -21.17 & 1.07 & 0.03 & 0.60 & 0.34 & H\\
NGC6946 & 5.5 & SAB(rs)cd  & 6 & -20.92 & 0.38 & 0.04 & 0.64 & 0.32 & H\\
IC342 & 3.0 & SAB(rs)cd  & 6 & -21.35 & 0.10 & 0.01 & 0.45 & 0.22 & H\\
\multicolumn{9}{l}{}\\
\hline
\end{tabular}
\end{center}
{\scriptsize{\bf Columns Explanation:} Col(1): Common Source Names; 
Col(2):  Distance to the source in units of Mpc are all taken directly from H97 where distances for objects closer than 40 Mpc are adopted from Tully \& Shaya (1984) and those farther than 40 Mpc were obtained using systemic velocities and assuming H$_0$ = 75 km s$^{-1}$ Mpc$^{-1}$;
Col(3):  Hubble Type;
Col(4):  Numerical Hubble type index as listed in Ho, Filippenko, \& Sargent (1997);
Col(5):  Total absolute B magnitude corrected for extinction, adopted from Ho, Filippenko, \& Sargent (1997);
Col(6):  [OIII] to H$_\beta$ ratio taken from Ho, Filippenko, \& Sargent (1997);  
Col(7):  [OI] to H$_\alpha$ ratio taken from Ho, Filippenko, \& Sargent (1997); 
Col(8):  [NII] to H$_\alpha$ ratio taken from Ho, Filippenko, \& Sargent (1997);
Col(9):  [SII] to H$_\alpha$ ratio taken from Ho, Filippenko, \& Sargent (1997);
Col(10):  Optical classification of the source; ``H'' signifies HII region ratios, ``T'' represents transitional spectra between LINERs and HII regions, and ``2'' indicates that broad permitted lines were not found in the optical spectrum.}
\end{table*}

%**********************************************************************

Figure 2 shows the standard optical line ratio diagnostic diagrams (Veilleux \& Osterbrock 1987) widely used to classify AGN for the entire H97 sample with our Spitzer sample highlighted.  We also show in Figure 2 the theoretical starburst limit line from Kewley et al. 2001.  This line represents the maximum line ratios possible from starburst photoionization models using the hardest possible starburst input ionizing radiation field.  Note that the majority of galaxies in our sample ($\sim$85-94\%, depending on the diagram) have optical line ratios well to the left of this line, indicating that the optical line ratios do not require the presence of {\it any} AGN contribution.  We point out that the optical spectroscopic classification of a galaxy in principle will depend on the measurement aperture size, particularly in sources with weak AGN surrounded by vigorous star formation activity.  However, the aperture size employed in the H97 optical spectral measurements ($\sim$2\arcsec$\times$4\arcsec) are in general some of the smallest available in the literature.  In addition, the sensitivity of the observations is considerably higher relative to many other surveys; the optical line ratios from H97 therefore provide the most sensitive optical classifications for weak AGN currently available. 

%**********************************************************************
\begin{figure*}[]
\begin{center}
\begin{tabular}{cc}
  \includegraphics[width=0.45\textwidth]{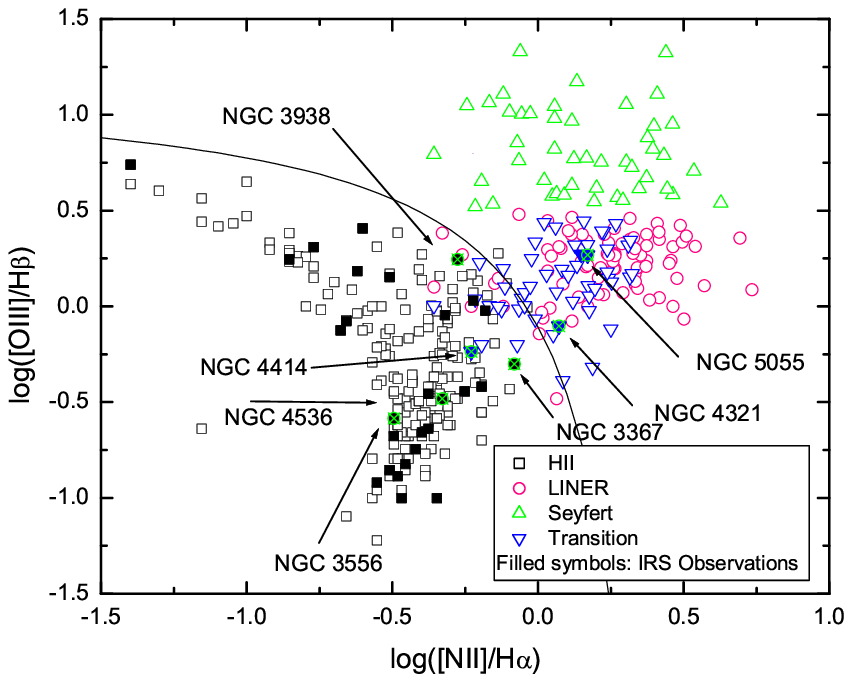} &
  \includegraphics[width=0.45\textwidth]{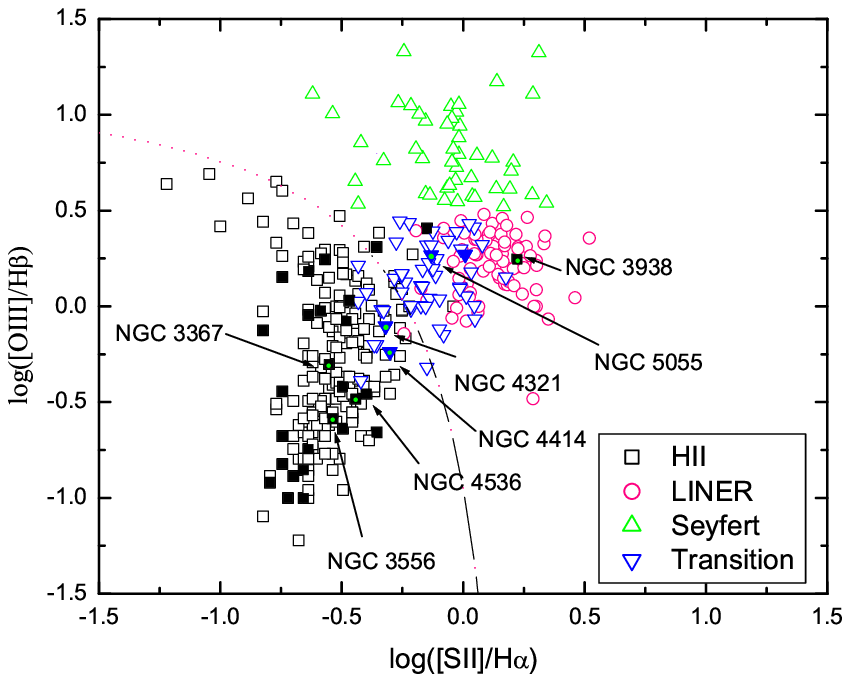} \\
  \multicolumn{2}{c}{\includegraphics[width=0.45\textwidth]{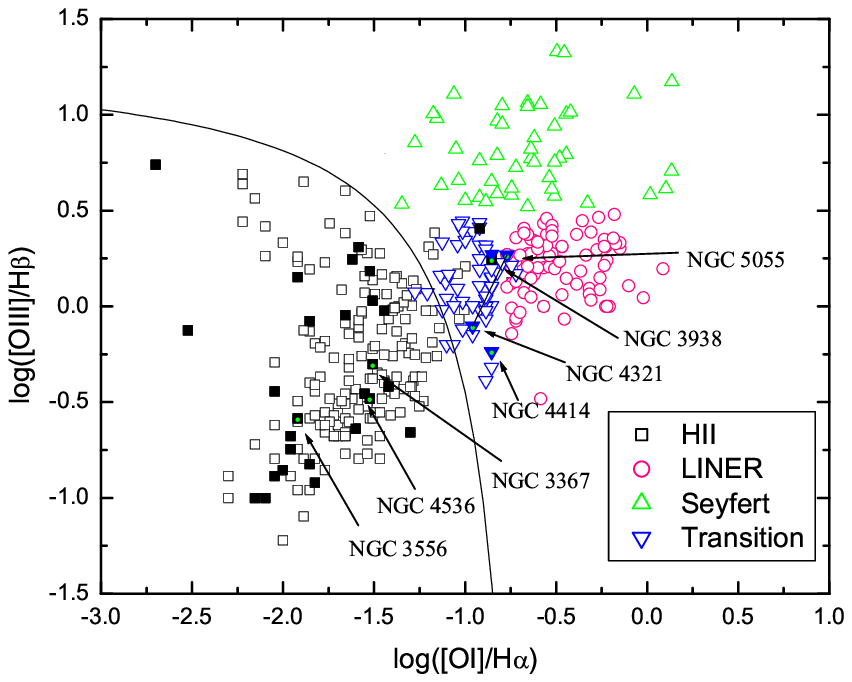}} \\
\end{tabular}
\end{center}
\caption[]{Standard optical line ratio diagnostic diagrams (Veilleux \& Osterbrock 1987) widely used to classify AGN for the entire H97 sample with our {\it Spitzer} sample represented by the filled symbols. The solid line represents the theoretical starburst limit line from Kewley et al. 2001.  This line represents the maximum line ratios possible from purely starburst photoionization models.  The galaxies with [NeV] detections are labeled in the figure (see Section 4 for details).    Note that most of the galaxies with [NeV] detections are to the left of the starburst limit line, indicating that their optical line ratios are consistent purely with star formation.}
\end{figure*}
%**********************************************************************

\section{Observational Details and Data Reduction}

All data presented in this work were obtained using both the short-wavelength (SH, 4.7''$\times$11.3'', $\lambda$  =  9.9-19.6$\mu$m) and long-wavelength(LH, 11.1''$\times$22.3'', $\lambda$ = 18.7-37.2$\mu$m) high-resolution modules of the IRS.  These modules have a spectral resolution of R $\sim$ 600.  The 32 observations were archived from various programs, including the {\it SINGS} Legacy Program (Kennicutt et al. 2003), and therefore contain both spectral mapping and staring observations.  In spectral mapping mode, the spacecraft moves in a raster pattern of discrete steps, generally of half-slit width and half-slit length in size, in order to construct a rectangular map of the targeted region.  The SH and LH maps for the observations vary in size.  In Table 2, we summarize the observational details for our sample.  The apertures from which our fluxes were extracted are listed in Table 2 and correspond to either the size of a single slit if the galaxy was observed as a staring observation, or the maximum overlap aperture between the SH and LH maps such that both the SH and LH spectra correspond to the same physical region on the sky.  For the {\it SINGS} observations, this corresponds to roughly a 23''$\times$15'' aperture.   In some mapping observations, the [NeV] line was detected in a small aperture, but not in the largest possible extraction region.  In these cases, the aperture used to obtain the [NeV] detection is listed in Table 2.  Here, identical apertures for both the SH and LH modules were also used so that spectra from both modules corresponded to the same physical region.
%**********************************************************************
\begin{table*}
\fontsize{8pt}{8pt}\selectfont
\begin{center}
\begin{tabular}{lclccrlcc}
\multicolumn{9}{l}{{\bf Table 2: Observational Details}}\\ 
\hline
\multicolumn{9}{l}{}\\
\multicolumn{1}{c}{Galaxy} & \multicolumn{1}{c}{PID} & \multicolumn{1}{c}{Instrument} & Exposure & Exposure & \multicolumn{2}{c}{Position (J2000)} & Extraction & \multicolumn{1}{c}{Extraction}\\

\multicolumn{1}{c}{Name} & & \multicolumn{1}{c}{Mode} & Time SH & Time LH & \multicolumn{1}{c}{RA} & \multicolumn{1}{c}{Dec} & Aperture & \multicolumn{1}{c}{Aperture}\\

\multicolumn{1}{c}{} & & \multicolumn{1}{c}{} & & & \multicolumn{1}{c}{} & \multicolumn{1}{c}{} & Size SH & \multicolumn{1}{c}{Size LH}\\

\multicolumn{1}{c}{(1)} & \multicolumn{1}{c}{(2)} & \multicolumn{1}{c}{(3)} & (4) & (5) & \multicolumn{1}{c}{(6)} & \multicolumn{1}{c}{(7)} & (8) & \multicolumn{1}{c}{(9)}\\ 
\multicolumn{9}{l}{}\\
\hline
\multicolumn{9}{l}{}\\
NGC925 & 159 & Mapping & 60 & 60 & 02:27:16.90 & +33:34:45.00 & 14$\times$25 & 14$\times$25\\
NGC1569  & 117 & Mapping & 600 & 1200 & 04:30:49.10 & +64:50:53.0 & 5$\times$57 & 22$\times$33\\
NGC2276 & 20140 & Staring & 60 & 120 & 07:27:14.30 & +85:45:16.00 & 5$\times$11 & 11$\times$22\\
NGC2903 & 59 & Staring & 120 & 120 & 09:32:10.1 & +21:30:04 & 5$\times$11 & 11$\times$22\\
NGC2976 & 159 & Mapping & 60 & 60 & 09:47:15.40 & +67:54:59.00 & 14$\times$25 & 14$\times$25\\
NGC3034 & 21 & Mapping & 24 & 30 & 09:55:52.20 & +69:40:47.00 & 27$\times$27 & 27$\times$27\\
NGC3077 & 73 & Staring & 30 & 14 & 10:03:19.10 & +68:44:02.00 & 5$\times$11 & 11$\times$22\\
NGC3184 & 159 & Mapping & 60 & 60 & 10:18:17.00 & +41:25:28.00 & 14$\times$25 & 14$\times$25\\
NGC3310 & 14 & Staring & 120 & 120 & 10:38:45.96 & +53:30:05.3 & 5$\times$11 & 11$\times$22\\
NGC3367 & 59 & Staring & 240 & 120 & 10:46:34.96 & +13:45:02.8 & 5$\times$11 & 11$\times$22\\
NGC3198 & 159 & Mapping & 60 & 60 & 10:19:55.05 & +45:32:59.50 & 14$\times$25 & 14$\times$25\\
NGC3556 & 14 & Staring & 120 & 120 & 11:11:30.97 & +55:40:26.8 & 5$\times$11 & 11$\times$22\\
NGC3726 & 3124 & Staring & 240 & 240 & 11:33:21.17 & +47:01:44.7 & 5$\times$11 & 11$\times$22\\
NGC3938 & 159 & Mapping & 60 & 60 & 11:52:50.3 & +44:07:15 & 5$\times$11 & 11$\times$22\\
NGC3949 & 3124 & Staring & 240 & 240 & 11:53:41.74 & +47:51:31.5 & 5$\times$11 & 11$\times$22\\
NGC4088 & 14 & Staring & 120 & 120 & 12:05:34.19 & +50:32:20.5 & 5$\times$11 & 11$\times$22\\
NGC4192 & 3237 & Staring & 480 & 300 & 12:13:48.29 & +14:54:01.2 & 5$\times$11 & 11$\times$22\\
NGC4236 & 159 & Mapping & 60 & 60 & 12:16:42.10 & +69:27:45.00 & 14$\times$25 & 14$\times$25\\
NGC4273 & 20140 & Staring & 60 & 120 & 12:19:56.1 & +05:20:35 & 5$\times$11 & 11$\times$22\\
NGC4254 & 159 & Mapping & 60 & 60 & 12:18:49.60 & +14:24:59.00 & 14$\times$25 & 14$\times$25\\
NGC4321 & 159 & Mapping & 60 & 60 & 12:22:54.93 & +15:49:21.8 & 27$\times$43 & 31$\times$49\\
NGC4490 & 3124 & Staring & 240 & 240 & 12:30:36.37 & +41:38:37.1 & 5$\times$11 & 11$\times$22\\
NGC4414 & 3674 & Staring & 240 & 180 & 12:26:27.10 & +31:13:24.7 & 5$\times$11 & 11$\times$22\\
NGC4536 & 159 & Mapping & 60 & 60 & 12:34:27.35 & +02:11:25.5 & 9$\times$9 & 9$\times$9\\
NGC4559 & 159 & Mapping & 60 & 60 & 12:35:57.70 & +27:57:35.00 & 14$\times$25 & 14$\times$25\\
NGC4567 & 20140 & Staring & 60 & 120 & 12:36:32.8 & +11:15:26 & 5$\times$11 & 11$\times$22\\
NGC4631 & 159 & Mapping & 60 & 60 & 12:42:08.01 & +32:32:29.4 & 16$\times$25 & 16$\times$25\\
NGC5055 & 159 & Mapping & 60 & 60 & 13:15:49.3 & +42:01:52 & 5$\times$11 & 11$\times$22\\
NGC5474 & 159 & Mapping & 60 & 60 & 14:05:01.61 & +53:39:44.00 & 14$\times$25 & 14$\times$25\\
NGC5907 & 3124 & Staring & 240 & 240 & 15:15:53.69 & +56:19:43.9 & 5$\times$11 & 11$\times$22\\
NGC6946 & 159 & Mapping & 60 & 60 & 20:34:52.30 & +60:09:14.00 & 18$\times$31 & 18$\times$31\\
IC342 & 14 & Staring & 120 & 112 & 3:46:48.51 & +68:05:46.0 & 5$\times$11 & 11$\times$22\\
\multicolumn{9}{l}{}\\
\hline
\end{tabular}
\end{center}
{\scriptsize{\bf Columns Explanation:} 
Col(1):  Common Source Names; 
Col(2):  Project identification number
Cols(3):  Instrument mode for the observation;
Col(4) \& (5):  Exposure time per pointing in seconds given for the SH and LH modules, respectively;
Col(6) \& (7):  Position of peak flux of [NeV] 14\micron\ when detected and 2MASS or radio nuclear coordinates otherwise;
Col(8) \& (9):  Extraction apertures for mapping observations could be matched using CUBISM, whereas single slit sizes are quoted for staring observations.  Values given are in arcseconds.}
\end{table*}
%**********************************************************************
The data presented here were preprocessed by the IRS pipeline (version 15.3) at the {\it Spitzer} Science Center (SSC) prior to download.  Preprocessing includes ramp fitting, dark-sky subtraction, droop correction, linearity correction, flat-fielding, and flux calibration\footnote[1]{See {\it Spitzer} Observers Manual, Chapter 7, http://ssc.spitzer.caltech.edu/documents/som/som8.0.irs.pdf}.  For all mapping observations, we used the pipeline-processed 'BCD-level' data products downloaded from the {\it Spitzer} archive in conjunction with \emph{CUBISM} v.1.5\footnote[2]{URL:  http:// ssc.spitzer.caltech.edu/ archanaly/ contributed/ cubism/ index.html } (Kennicutt et al. 2003; Smith et al. 2004) to construct the high-resolution spectral cubes. The 2D FITS output images from {\it CUBISM} were smoothed using a Gaussian kernel of 1.5-pixel width.  To extract final 1D spectra and line maps, we used the {\it CUBEVIEW} and {\it CUBESPEC} tools within \emph{CUBISM}. Detailed descriptions of the post-processing steps included in {\it CUBISM} are given in Smith et al. (2004).  The {\it Spitzer} spectra obtained from both staring and mapping observations were further processed using the SMART v. 6.2.4 analysis package (Higdon et al. 2004) and the corresponding version of the calibration files (v.1.4.8), which were used to obtain final line fluxes.  

%**********************************************************************
\begin{table}
\fontsize{8pt}{9pt}\selectfont
\begin{center}
\begin{tabular}{lccc}
\multicolumn{4}{l}{{\bf Table 3: Fluxes for Full Sample}}\\ 
\hline

\multicolumn{4}{l}{}\\
\multicolumn{1}{c}{Galaxy} & [NeV] 14.32\micron & [NeV] 24.32\micron & [OIV] 25.89\micron\\
\multicolumn{1}{c}{Name} & (97.1 eV) & (97.1 eV) & (54.9 eV)\\

\multicolumn{1}{c}{(1)} & (2) & \multicolumn{1}{c}{(3)} & (4)\\ 
\multicolumn{4}{l}{}\\
\hline
\multicolumn{4}{l}{}\\
NGC925 & $<$1.0 & $<$0.48 & 0.63$\pm$0.19\\
NGC1569  & $<$0.15 & $<$1.0 & 25.0$\pm$6.6\\
NGC2276 & $<$0.56 & $<$0.39 & $<$0.69\\
NGC2903 & $<$0.47 & $<$2.7 & $<$4.0\\
NGC2976 & $<$0.81 & $<$0.37 & $<$0.28\\
NGC3034 & $<$111.1 & $<$250.6 & $<$259.6\\
NGC3077 & $<$0.45 & $<$0.75 & $<$0.81\\
NGC3184 & $<$0.70 & $<$0.59 & 0.57$\pm$0.17\\
NGC3310 & $<$0.38 & $<$1.7 & 3.9$\pm$0.8\\
NGC3367 & 1.2$\pm$0.3 & 0.93$\pm$0.24 & 0.81$\pm$0.23\\
NGC3198 & $<$0.80 & $<$0.93 & $<$0.82\\
NGC3556 & 0.40$\pm$0.09 & $<$1.1 & $<$1.5\\
NGC3726 & $<$0.30 & $<$0.66 & $<$0.62\\
NGC3938 & $<$4.43 & 1.1$\pm$0.3 & 0.56$\pm$0.19\\
NGC3949 & $<$0.31 & $<$0.54 & 0.92$\pm$0.29\\
NGC4088 & $<$0.39 & $<$0.68 & $<$0.73\\
NGC4192 & $<$0.59 & $<$1.5 & $<$1.2\\
NGC4236 & $<$0.85 & $<$0.45 & $<$0.41\\
NGC4273 & $<$0.62 & $<$0.40 & $<$1.28\\
NGC4254 & $<$0.90 & $<$0.41 & 1.6$\pm$0.5\\
NGC4321 & 2.6$\pm$0.8 & $<$3.6 & $<$5.6\\
NGC4490 & $<$0.22 & $<$0.46 & 1.3$\pm$0.2\\
NGC4414 & $<$0.65 & 1.2$\pm$0.4 & 2.1$\pm$0.3\\
NGC4536 & 0.32$\pm$0.09 & $<$0.90 & $<$1.2\\
NGC4559 & $<$0.84 & $<$0.39 & $<$0.40\\
NGC4567 & $<$0.53 & $<$0.45 & $<$0.54\\
NGC4631 & $<$2.0 & $<$1.35 & $<$4.7\\
NGC5055 & 0.56$\pm$0.18 & $<$0.55 & 1.4$\pm$0.2\\
NGC5474 & $<$1.1 & $<$0.44 & $<$0.32\\
NGC5907 & $<$0.23 & $<$0.58 & 1.3$\pm$0.2\\
NGC6946 & $<$1.5 & $<$0.39 & $<$12.07\\
IC342 & $<$1.7 & $<$15.88 & $<$23.37\\
\multicolumn{4}{l}{}\\
\hline
\end{tabular}
\end{center}
{\scriptsize{\bf Columns Explanation:} 
Col(1): Common Source Names; 
Col(2) \& (3) \& (4):  Fluxes are in units of 10$^{-21}$ W cm$^{-2}$. 3 $\sigma$ upper limits are reported for nondetections.}
\end{table}
%**********************************************************************
All of the staring observations were centered on the nucleus of the galaxy as requested, and generally correspond to the {\it 2MASS} nuclear coordinates..  We note that the SH and LH staring observations include data from two slit positions overlapping by one third of a slit.   Because the slits occupy distinctly different regions of the sky, the flux from the slits cannot be averaged unless the emission originates from a compact source that is contained entirely in each slit.  Therefore, the procedure for flux extraction for staring observations was the following: 1) If the fluxes measured from the two slits differed by no more than the calibration error of the instrument, then the fluxes were averaged; otherwise, the slit with the highest measured line flux was chosen.  2) If an emission line was detected in one slit, but not in the other, then the detection was selected.  The slit for the SH and LH modules is too small for background subtraction to take place and separate SH or LH background observations do not exist for any of the galaxies in this sample.  We therefore did not perform any background subtraction on any of the observations presented in this work.  The overall calibration uncertainty for the fluxes we report in this paper is 25 to 30\% for all mapping observations and 15\% for staring observations.  

\section{Results}

\subsection{ Line Fluxes and Spectral Line Fits }

In Table 3 we list the line fluxes, statistical errors and upper limits for the SH and LH observations for the [NeV] 14.3$\mu$m and 24.3$\mu$m lines, as well as the 26$\mu$m [OIV] line.  The apertures from which these fluxes were extracted are listed in Table 2.  In all cases, detections were defined when the line flux was at least 3$\sigma$.  We detected either the [NeV] 14.3$\mu$m or 24.3$\mu$m line in 7 out of the 32 galaxies, providing strong evidence for the presence of an AGN in these galaxies..  The [OIV] 26$\mu$m line (ionization potential = 55 eV) was detected in 12 out of the 32 galaxies.  Although this line is generally strong in powerful AGN (e.g. Genzel et al. 1998, Sturm et al. 2002), it can be produced purely by photoionization by very hot stars or shock-heated gas (Schaerer \& Stasinska 1999) and is therefore not a definitive signature of an AGN.  The spectral line fits for the detected [NeV] lines as well as the [OIV] 26$\mu$m and [NeIII] 15.5$\mu$m lines, when detected, for each of the 7 galaxies are shown in Figure 3. We note that the spectral resolution of the SH and LH modules of the IRS is insufficient to resolve the velocity structure for most of the lines.  

The optical line ratios of the galaxies with [NeV] detections are shown in Figure 1.  As can be seen, several of the galaxies with [NeV] detections lie significantly to the left of the theoretical limit line from Kewley et al. (2001) indicating that there is no hint of the presence of an AGN from the optical emission lines alone.  Three out of the 7 galaxies with [NeV] detections are classified by H97 as transition objects as can be seen in Table 1.  The [NeV] 14.3$\mu$m or 24.3$\mu$m line was not detected in only one out of the 4 transition objects in our sample.  

\subsection{ IR Emission Line Morphologies}

The signal-to-noise of the observations was insufficient to provide detailed spatial information on the [NeV] emission for all 4 of the AGN candidates with mapping observations.  However, we show in Figure 4, the detected [OIV] 26$\mu$m  and [NeIII] 15.5$\mu$m images for 3 out of the 4 galaxies with [NeV] detections.  The signal-to-noise ratio corresponding to both the [OIV] and the [NeIII] for NGC3968 precluded any analysis of the spatial distribution of the emission in this galaxy.  This galaxy is therefore excluded from Figure 4.  The ionization potential of OIV and NeIII are 55 eV and 41 eV, respectively.  Both species can be produced in gas ionized by hot stars. However in all cases, centrally compact emission is observed, suggestive of an AGN origin..  Although we do not know the spatial morphology of the [NeV] emission, it is likely to be similar to the centrally concentrated [OIV] and [NeIII] emission shown below.  We note that there is possibly one anomalous galaxy in our sample of detections.  The peak of the [NeV] emission in NGC 4536 appears to be offset from the peak of the [OIV] and [NeIII] emission which both peak at the optical and {\it 2MASS} nucleus.  However, the signal to noise ratio of the data severely hampers any detailed study of the spatial distribution of the [NeV] emission in this galaxy (see Section 5.1 for details).  We discuss each galaxy individually in Section 5.1.

\subsection{ Line Flux Ratios and Luminosities}

Since the ratio of high to low excitation lines depends on the nature of the ionizing source, the [NeV]14$\mu$m/[NeII]12.8$\mu$m and the  [OIV]25.9$\mu$m/[NeII]12.8$\mu$m line flux ratios have been used to characterize the nature of the dominant ionizing source in galaxies (Genzel et al. 1998; Sturm et al. 2002; Satyapal et al. 2004; Dale et al. 2006). Table 4 lists the [NeV]/[NeII], [NeV]/[NeIII], and [OIV]/[NeII] line flux ratios and the [NeV] 14.3$\mu$m line luminosities  for the 7 galaxies in which either [NeV] line is detected.  In NGC 3938 and NGC 4414 the [NeV] 14.3$\mu$m line was not detected.  In this case, we estimated and use in Table 4 the 14.3$\mu$m line luminosity using the [NeV] 24.3$\mu$m line luminosity.  The [NeV] 14.3$\mu$m and 24.3$\mu$m line luminosities are tightly correlated in a large sample of standard optically-identified AGN recently observed by {\it Spitzer} (Dudik et al. 2007) as can be seen in Figure 5.  Employing a Spearman rank correlation analysis (Kendall \& Stuart 1976) yields a correlation coefficient of r$_{\rm S}$ = 0.97 between the [NeV] 14.3$\mu$m and 24.3$\mu$m line luminosities, with a probability of chance correlation of 6.7 $\times$ 10$^{-12}$, indicating a significant correlation.  This relation implies similar densities to the [NeV]-emitting gas in the galaxies in the sample.
The best-fit linear relation yields:
\begin{equation}
\log(L_{\rm [NeV]14}) =  (0.995)\log(L_{\rm [NeV]24}) + 0.0566
\end{equation}

with an rms scatter of 0.219 dex.  We note that equation (1) is consistent with the relationship between the [NeV] 14.3$\mu$m and 24.3$\mu$m line luminosities in standard AGN observed by {\it ISO} (Sturm et al. 2002), as well as NGC 3367, the one galaxy in our current sample in which both lines were detected.  The [NeV]/[NeII] line flux ratio for the 13 AGN with both [NeV] and [NeII] detections by the {\it Infrared Space Observatory (ISO)} ranges from 0.06 to 2.11, with a median value of 0.47 (Sturm et al. 2002). [NeV]/[NeII] ratios for our 7 galaxies range from approximately 0.01 to 0.27, with most falling below the lowest value observed in standard AGN by ISO.  The [OIV]/[NeII] line flux ratio for the 17 {\it ISO} AGN with both lines detected ranges from 0.15 to 8.33, with a median value of 1.73 (Sturm et al. 2002).  The [OIV]/[NeII] line flux ratio for our sample of 7 galaxies range from 0.28 to 0.67, within the range observed in the nearby powerful AGN observed by ISO. For comparison, the few starburst galaxies that show detectable [OIV] emission have [OIV]/[NeII] line flux ratios that range from 0.006 to 0.647 (median = 0.019; Verma et al. 2003) but show no [NeV] emission. 

We note that when mapping observations were used, line fluxes from SH and LH maps were extracted from identically sized apertures.  However, because the LH slit is larger than the SH slit, fluxes obtained from the SH and LH modules for staring observations correspond to different extraction regions.  The line ratios in this work as well as those obtained from {\it ISO} are subject to aperture effects and should be viewed with some caution.  In particular, the emission from lower ionization potential species are likely to be more spatially extended than the emission from higher ionization potential species.  Line flux ratios from both matched and non-uniform apertures with the SH and LH modules will therefore likely depend on the distance to the source.

Table 4 lists the [NeV] 14$\mu$m  line luminosities for all 7 of our galaxies. Compiling the line luminosities for the large (33) sample of optically identified standard AGN with [NeV] 14$\mu$m detections recently observed by {\it Spitzer} from Dudik et al. (2007) and Gorijian et al. (2007) , the [NeV] 14$\mu$m  line luminosity ranges from $\sim$ 1$\times$10$^{38}$ ergs s$^{-1}$ to $\sim$ 8$\times$10$^{42}$ ergs s$^{-1}$ with a median value of $\sim$ 1$\times$10$^{41}$ ergs s$^{-1}$.  As expected for weak AGN, the line luminosities for our candidate sample of AGN is significantly lower than that found for optically identified AGN.  Four out of the 7 galaxies with [NeV] detections have luminosities below the minimum value found in standard optically identified AGN.  Although NGC 4321, optically identified as a transition object, and NGC 3367 both have [NeV] luminosities somewhat higher than the minimum value seen in standard AGN, they are still both an order of magnitude less luminous than the median value observed in optically identified AGN.

In Figure 6, we investigate the relationship between the [OIII] $\lambda$5007 luminosity, for which extensive compilations are available in the literature, and the [NeV] luminosity in our sample of galaxies compared with standard AGN.  The [OIII] $\lambda$5007 emission is often assumed to originate primarily in gas ionized exclusively by the AGN and lying outside of the obscuring torus.  Under these assumptions, it has been used as a measure of the intrinsic bolometric luminosity of the AGN (e.g. Heckman et al. 2004).  As can be seen from Figure 6, there is a significant correlation between the [OIII] and [NeV] luminosities.  Employing a Spearman rank correlation analysis for the standard AGN plotted in Figure 6 yields a correlation coefficient between the line luminosities of r$_{\rm S}$ = 0.84, with a probability of chance correlation of 8 $\times$ 10$^{-6}$.  The best-fit linear regression yields the following relation:
\begin{equation}
\log(L_{\rm [OIII]\lambda5007}) =  (1.273)\log(L_{\rm [NeV]14}) - 10.268
\end{equation}

 The [NeV] data for the "standard AGN" plotted in Figure 6 were compiled from Dudik et al. (2007), Weedman et al. (2005), Ogle et al. (2006), Gorjian et al. (2007), Haas et al. (2005), Cleary et al. (2007), and Armus et al. (2007).  The "standard" AGN  are mostly Seyfert 1 galaxies, quasars, 3C radio galaxies, and a few bonafide Seyfert 2s but all LINERs were excluded.  The [OIII] luminosities were taken from H97, Veilleux et al. (1997), and the extensive compilations of [OIII] $\lambda$5007 emission line data from Xu et al. (1999) and Whittle (1992).  We note that none of the line luminosities plotted in Figure 6 are corrected for extinction.

There is considerable scatter in Figure 6 (rms scatter of 0.67 dex).  This scatter is expected for a number of reasons.  Firstly, the [OIII] fluxes used in Figure 6 were obtained using a wide range of apertures.  Since [OIII] emission can originate in both HII regions and the AGN narrow line region, the luminosity can vary considerably with measurement aperture size and the distance to the galaxy.  For example, [OIII] $\lambda$5007 measurements were recently obtained by Moustakas \& Kennicutt (2006) for two of the galaxies in our sample (NGC~4321 and NGC~4414).  These observations show that the [OIII] $\lambda$5007 nuclear (2.5''$\times$2.5'') flux is roughly two orders of magnitude lower than the flux integrated across the optical extent of the galaxy.  Thus, the AGN's contribution to the total luminosity is variable in the sample. Since the high ionization [NeV] line is more sensitive to the AGN, having a negligible star formation component, variation in the [OIII]/[NeV] luminosity ratio with AGN strength is expected.  Secondly, the [OIII] luminosity is not corrected for extinction, and although the AGN contribution to the [OIII] originates outside the torus, there still can be some extinction from the host galaxy, which is probably higher in our of galaxies sample since they are late-type spirals.  Figure 6 shows that the median [OIII]/[NeV] luminosity ratio in our sample of late-type galaxies is roughly 20 times less than the median value of the ratio for the standard AGN.  Assuming a foreground screen geometry for the obscuring material, this difference in line flux ratio can be caused by an extinction of only A$_{V}$ $\sim$ 3 magnitudes.

\subsection{ IR Spectroscopic AGN Detection Rate }

Using IR spectroscopy to identify AGN, counting only the 7 firm [NeV] detections, the detection rate of AGN in our sample of optically normal late-type galaxies is $\sim$ 20\%.  However, the limited and variable signal-to-noise of the observations leaves open the possibility that the detection rate could be higher if the sensitivity of the observations were uniform and higher.  Figure 7 shows the distribution of detections as a function of the 3$\sigma$ [NeV] 14$\mu$m  line sensitivity.  Here galaxies are included in each bin if the 3$\sigma$ line sensitivity is equal to the value listed on the X-axis or better.  Although our sample size is limited, precluding us from making statistically firm conclusions, Figure 7 suggests that the detection rate is not likely to be greater than $\sim$30\% even if deeper exposures were obtained in all the observations.

From the H97 sample, out of the full sample of 486 galaxies, 207  are of Hubble type Sbc and later  and only 16 ($\sim$ 8\%) are optically classified as AGN.  Our IR spectroscopic technique suggests that the AGN detection rate in optically normal late-type galaxies is $\sim$ 30\% -- implying that the incidence of AGN in late-type galaxies is possibly greater than a factor of four higher than what  optical observations alone suggest.

Amongst the 7 AGN candidates in our sample, 3 are Sbc galaxies, 3 are Sc galaxies, and 1 is of Hubble type Scd.  Figure 1 shows the distribution of Hubble types for the sample with the 7 AGN candidates displayed by the shaded histogram.  Since the sensitivity of the observations varied across the sample, we also indicate with a downward arrow in Figure 1 the number of galaxies with [NeV] 14$\mu$m 3$\sigma$ line sensitivity of 10$^{38}$ ergs s$^{-1}$  or better.  Although the sample size is too small to make statistically firm statements, we note that none of the galaxies of Hubble type later than Scd display a [NeV] line.  Unfortunately there are only 3 galaxies of Hubble type Sd.  With such limited statistics it is not possible to say how common it is for Sd galaxies to host AGN.  A more extensive mid-infrared spectroscopic survey of Sd galaxies is crucial to obtain meaningful statistics on the number of completely bulgeless galaxies that contain accreting SMBHs.

%**********************************************************************
\begin{figure*}[htbp]
\begin{center}
\begin{tabular}{cc}
  \includegraphics[width=0.45\textwidth]{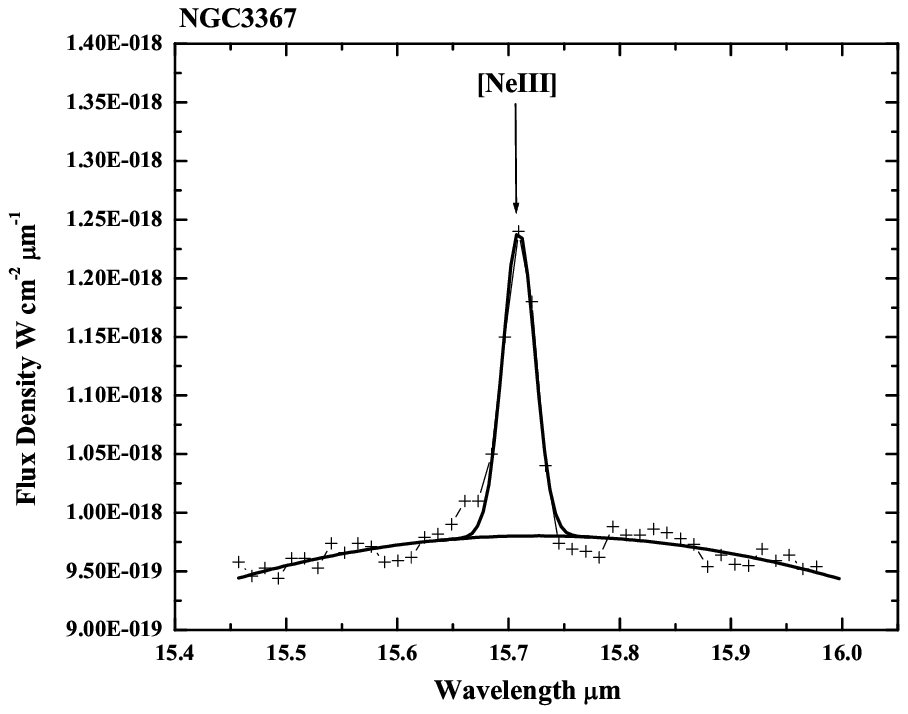} &
  \includegraphics[width=0.45\textwidth]{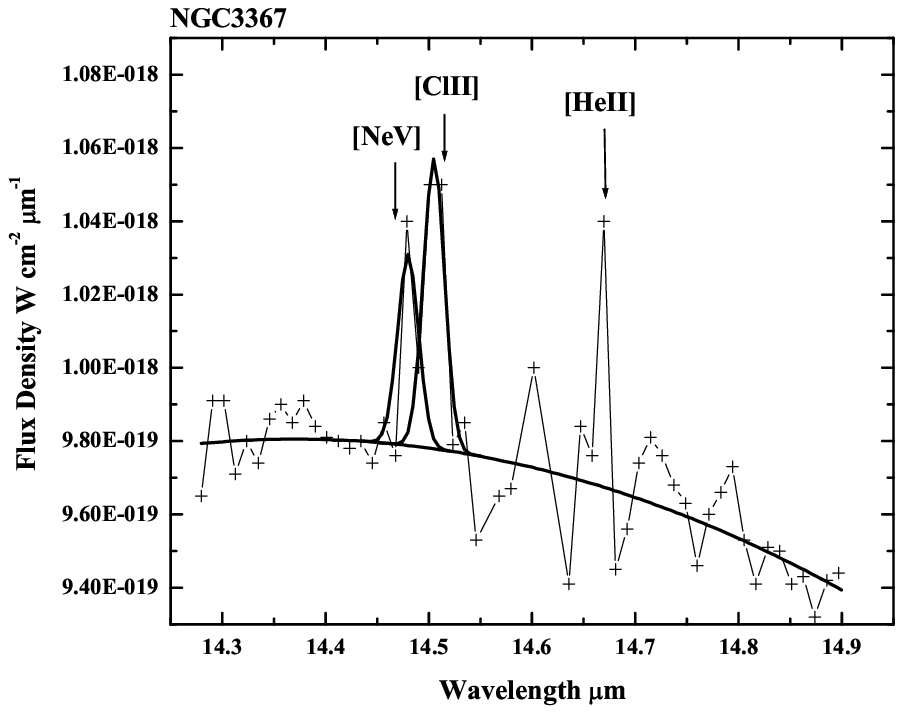}\\
  \includegraphics[width=0.45\textwidth]{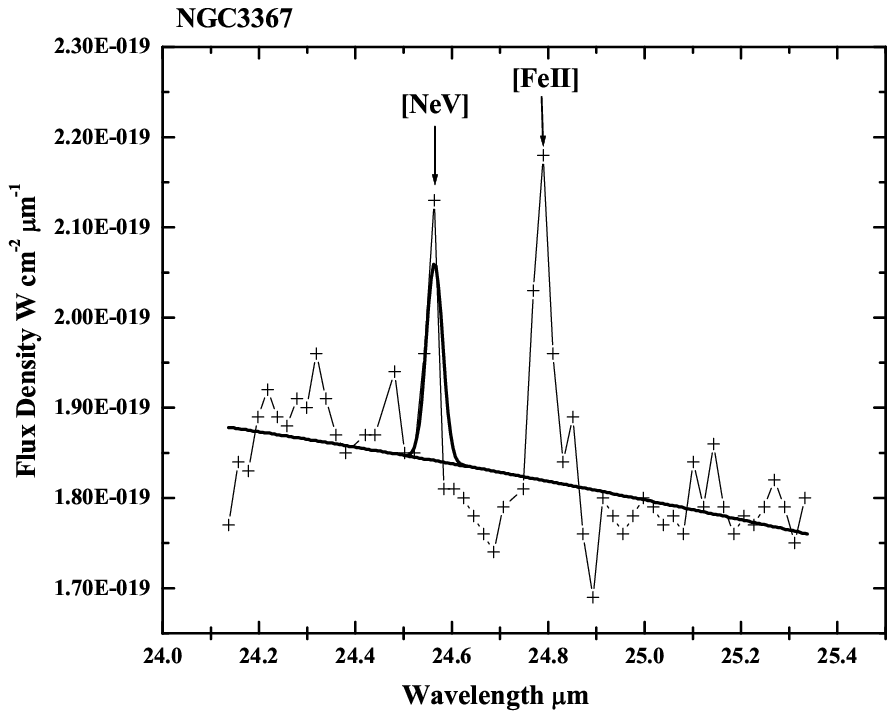} &
  \includegraphics[width=0.45\textwidth]{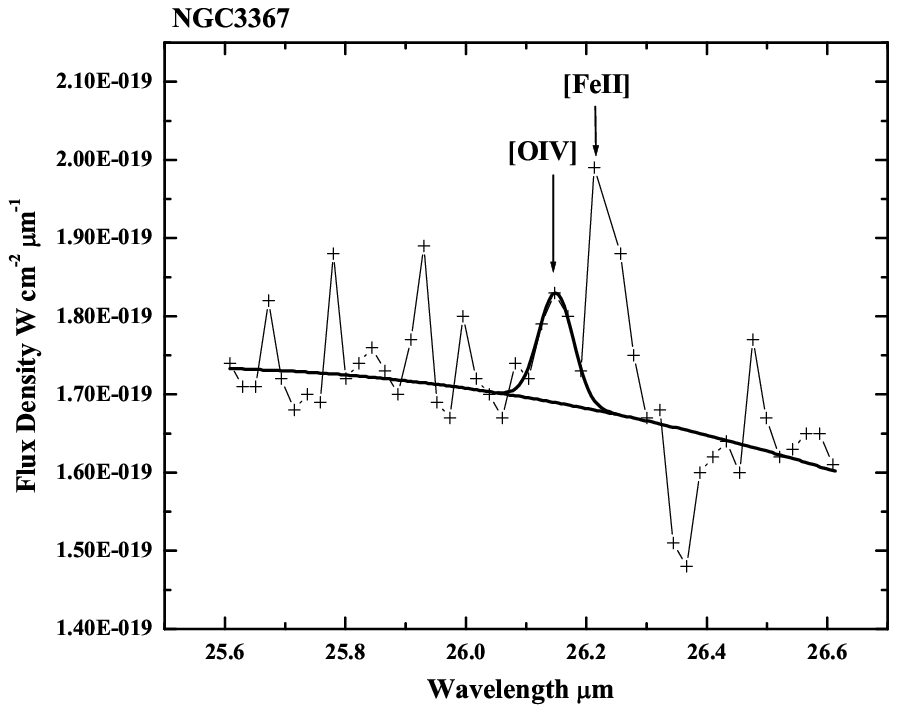} \\
  \includegraphics[width=0.45\textwidth]{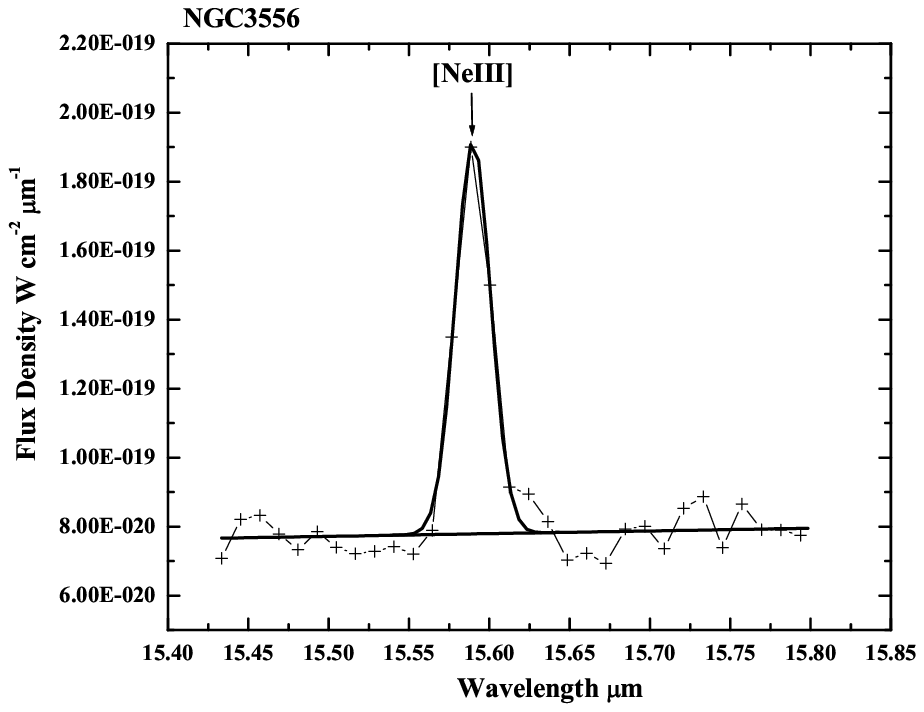} &
  \includegraphics[width=0.45\textwidth]{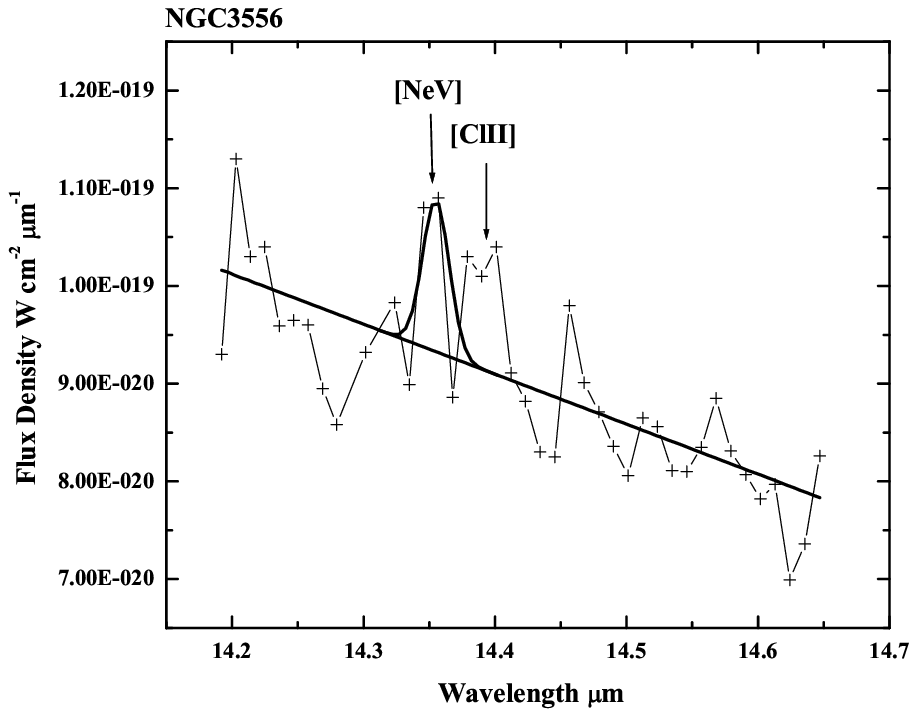} \\
\end{tabular}
\end{center}
\caption[]{IRS Spectra showing the detections of the [NeV] 14.3$\mu$m and/or [NeV] 24.3$\mu$m line for the 7 galaxies with detections listed in Table 3.  Also shown are spectra showing the [NeIII] 15.5$\mu$m and [OIV] 26$\mu$m lines when detected for each of the galaxies.
}
\end{figure*}
\begin{figure*}
\begin{center}
\begin{tabular}{cc}
  \includegraphics[width=0.45\textwidth]{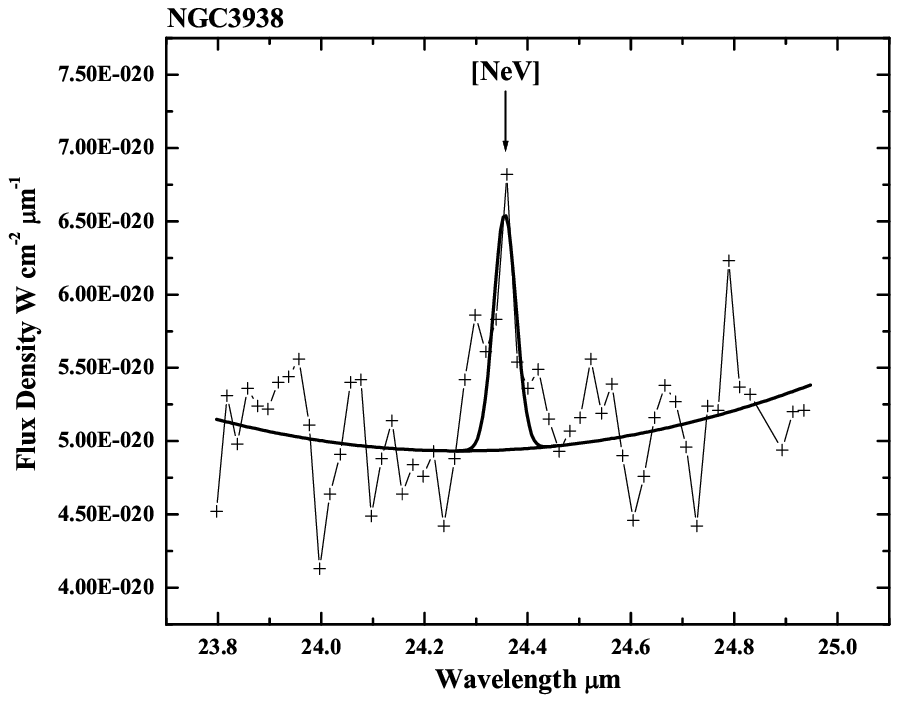} &
  \includegraphics[width=0.45\textwidth]{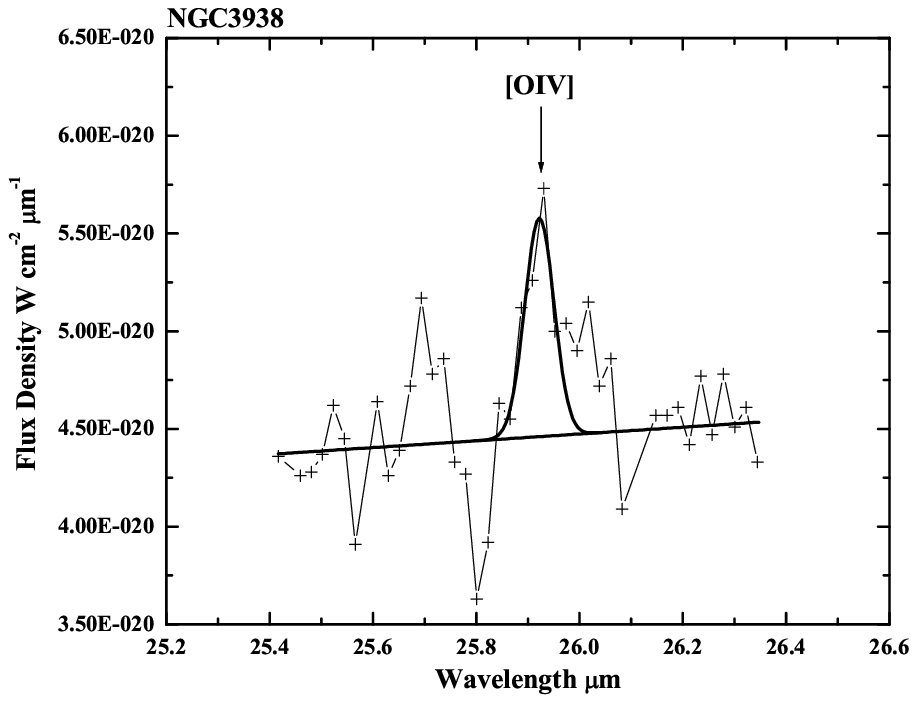}\\
  \includegraphics[width=0.45\textwidth]{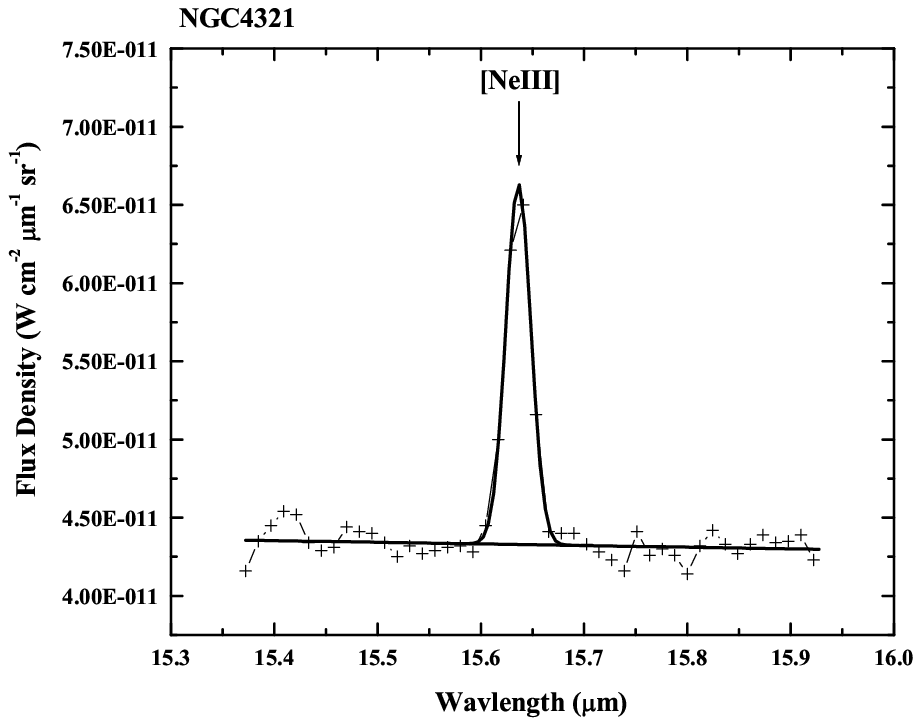}&
  \includegraphics[width=0.45\textwidth]{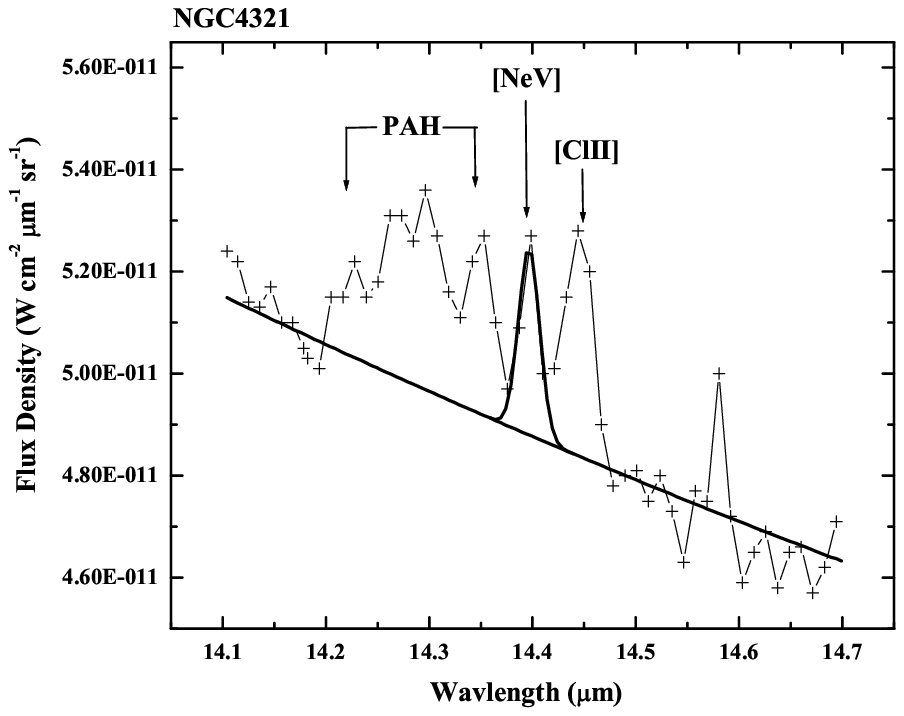}\\
 \includegraphics[width=0.45\textwidth]{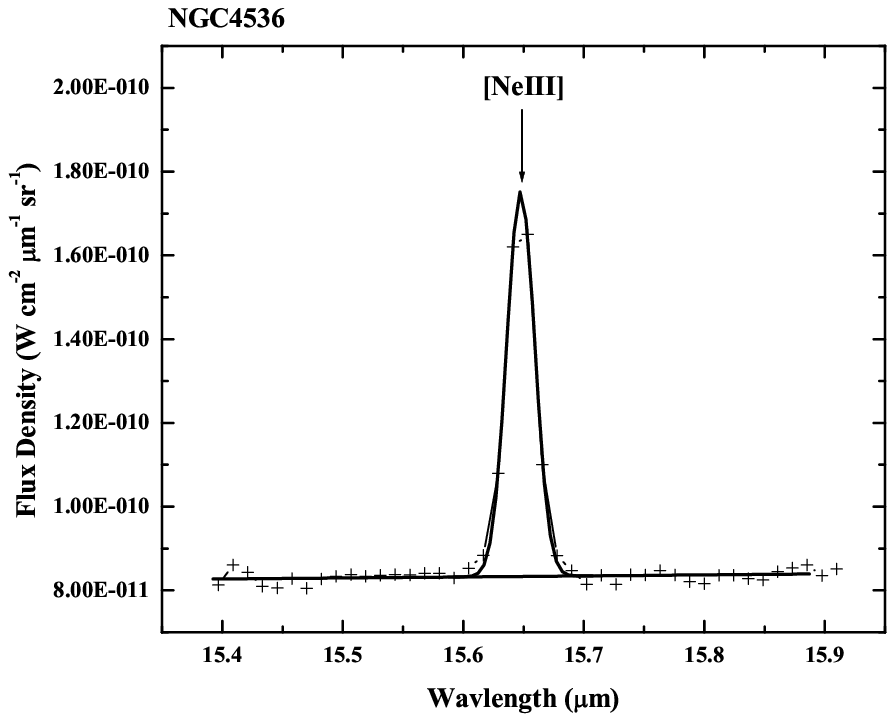} &
  \includegraphics[width=0.45\textwidth]{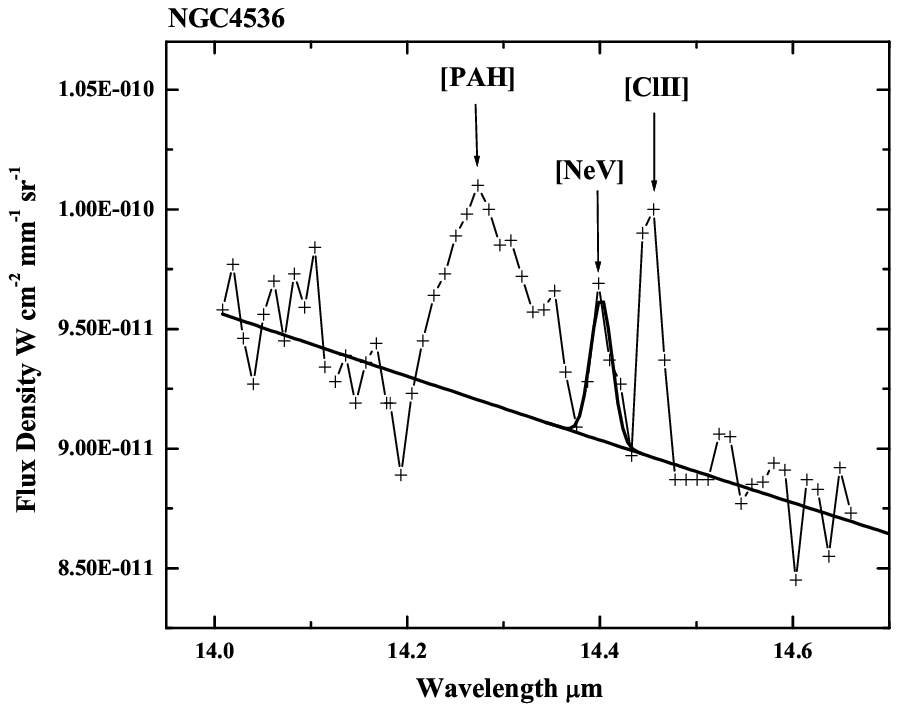} \\

  \multicolumn{2}{c}{\sc{Fig} 3.--{\it Continued}}\\
\end{tabular}
\end{center}
\end{figure*}

\begin{figure*}
\begin{center}
\begin{tabular}{cc}
   \includegraphics[width=0.45\textwidth]{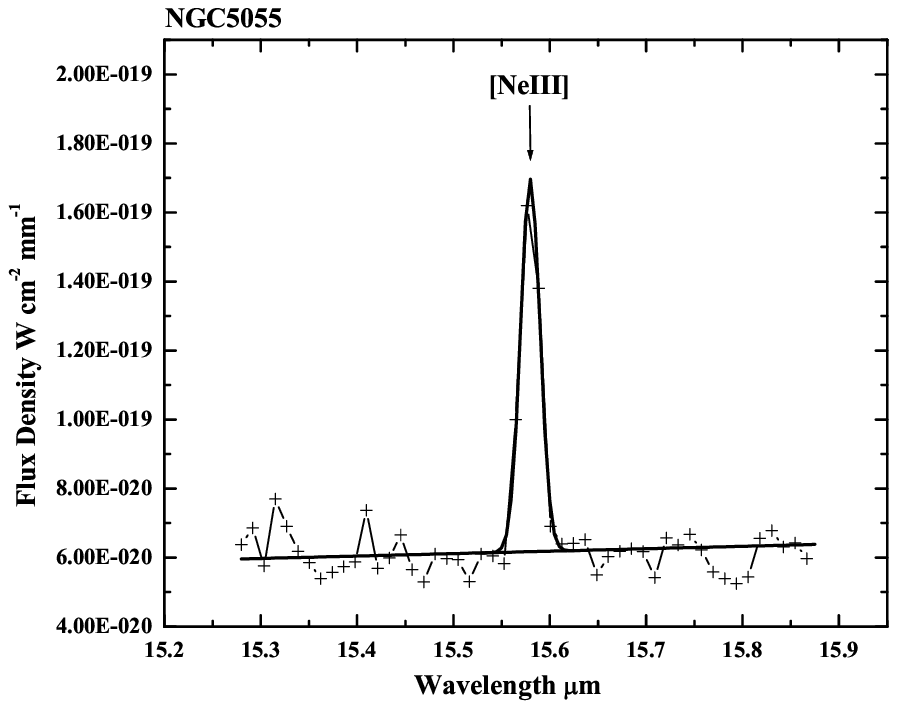} &
  \includegraphics[width=0.45\textwidth]{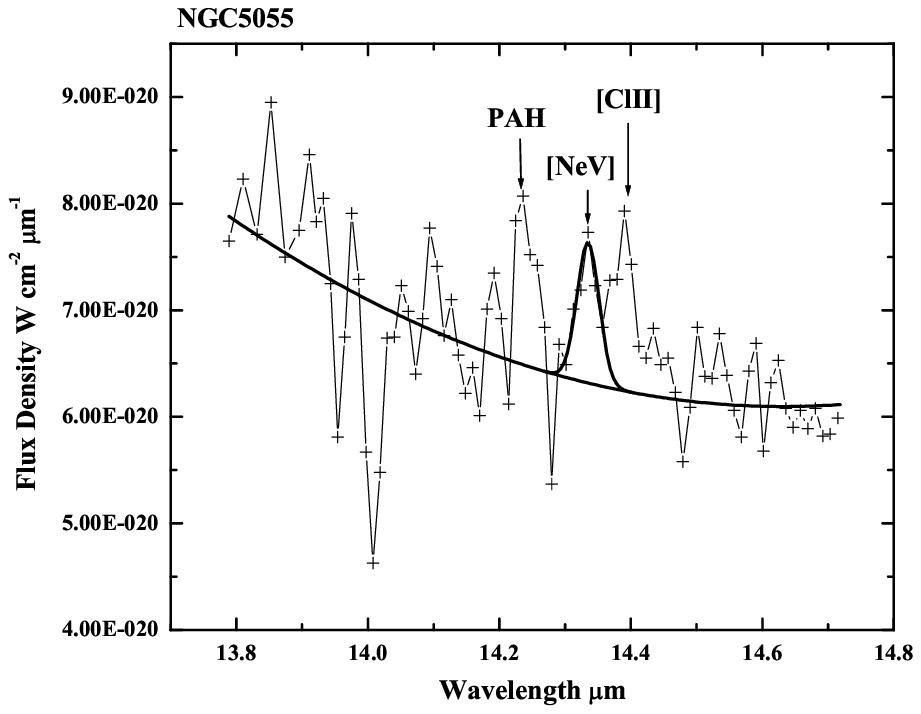} \\
 \includegraphics[width=0.45\textwidth]{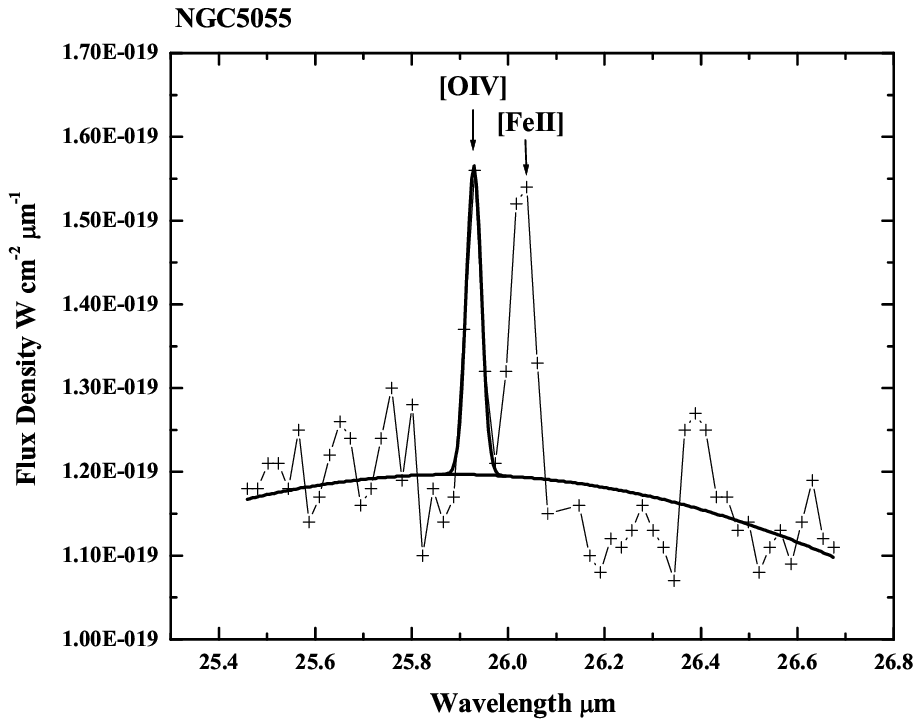} &
  \includegraphics[width=0.45\textwidth]{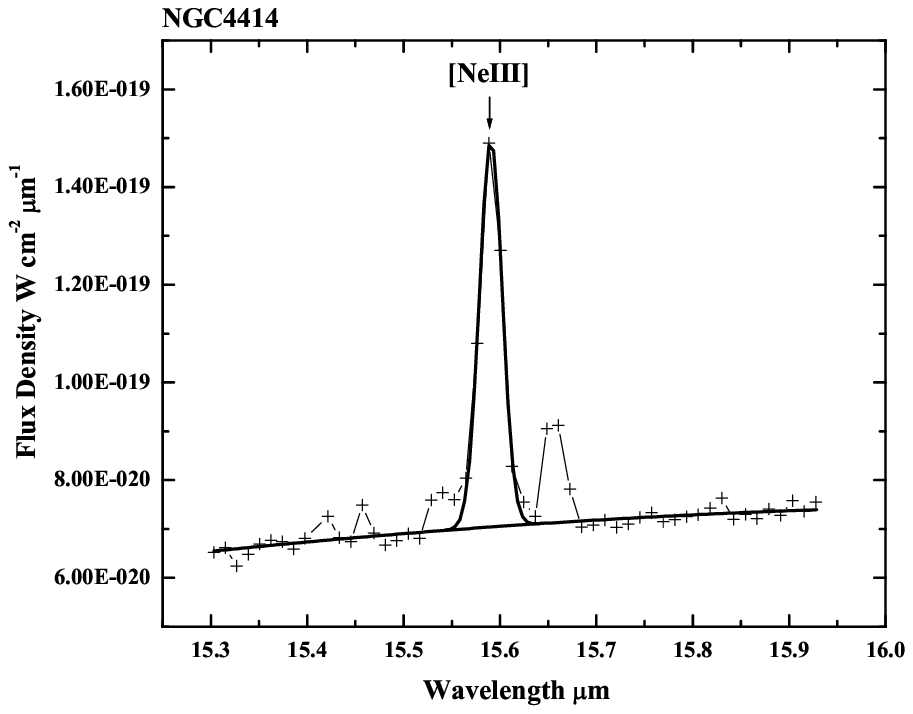} \\
  \includegraphics[width=0.45\textwidth]{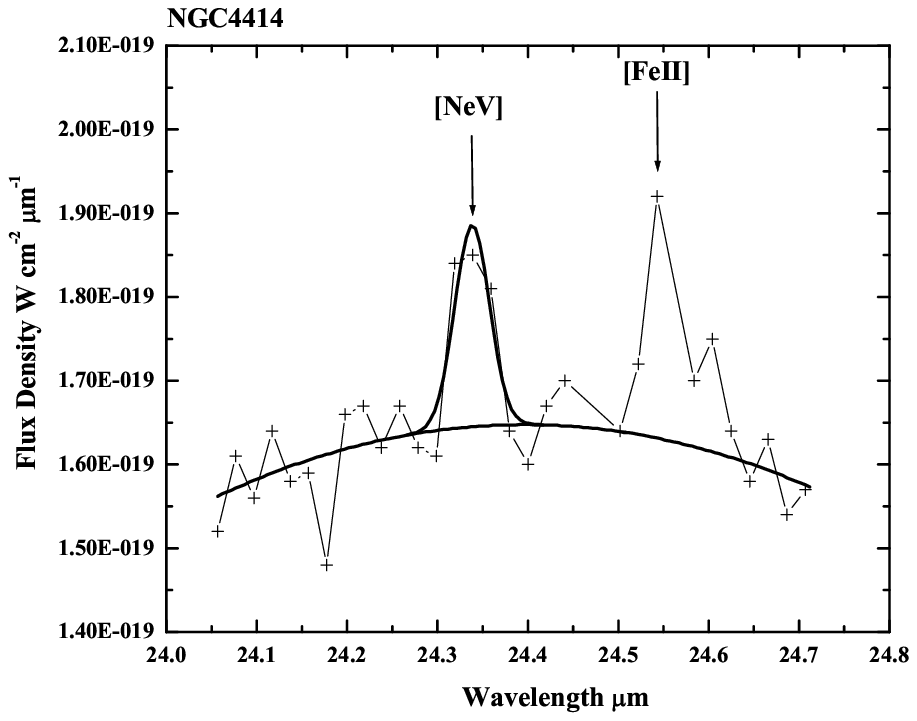} &
  \includegraphics[width=0.45\textwidth]{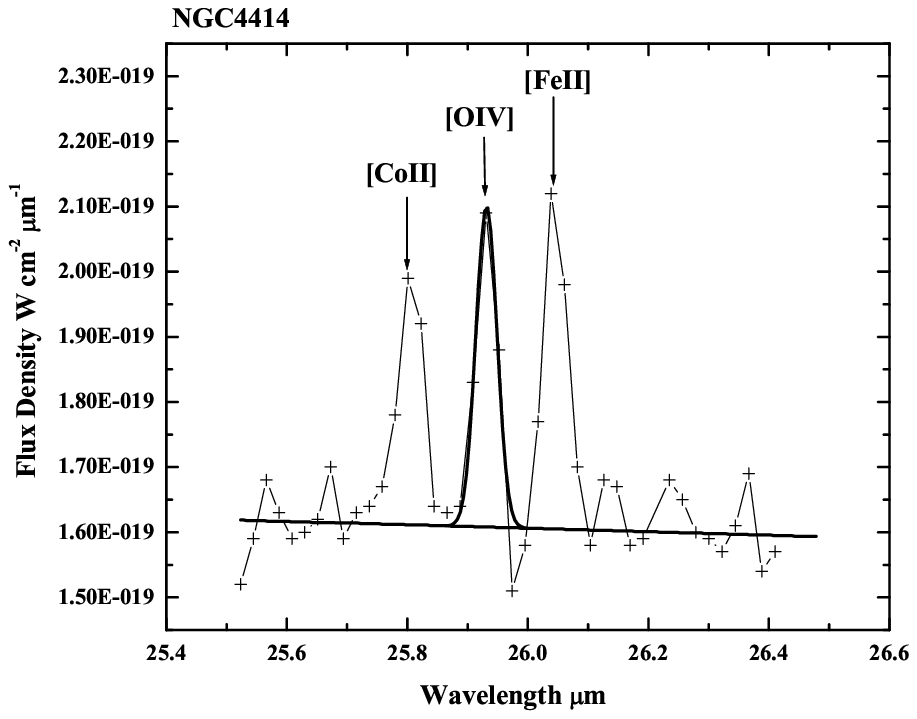} \\
   \multicolumn{2}{c}{{\sc Fig} 3.--{\it Continued}}\\
\end{tabular}
\end{center}
\end{figure*}
%**********************************************************************
%**********************************************************************
\begin{figure*}[]
\begin{center}
\begin{tabular}{cc}
  \includegraphics[width=0.38\textwidth,angle=0]{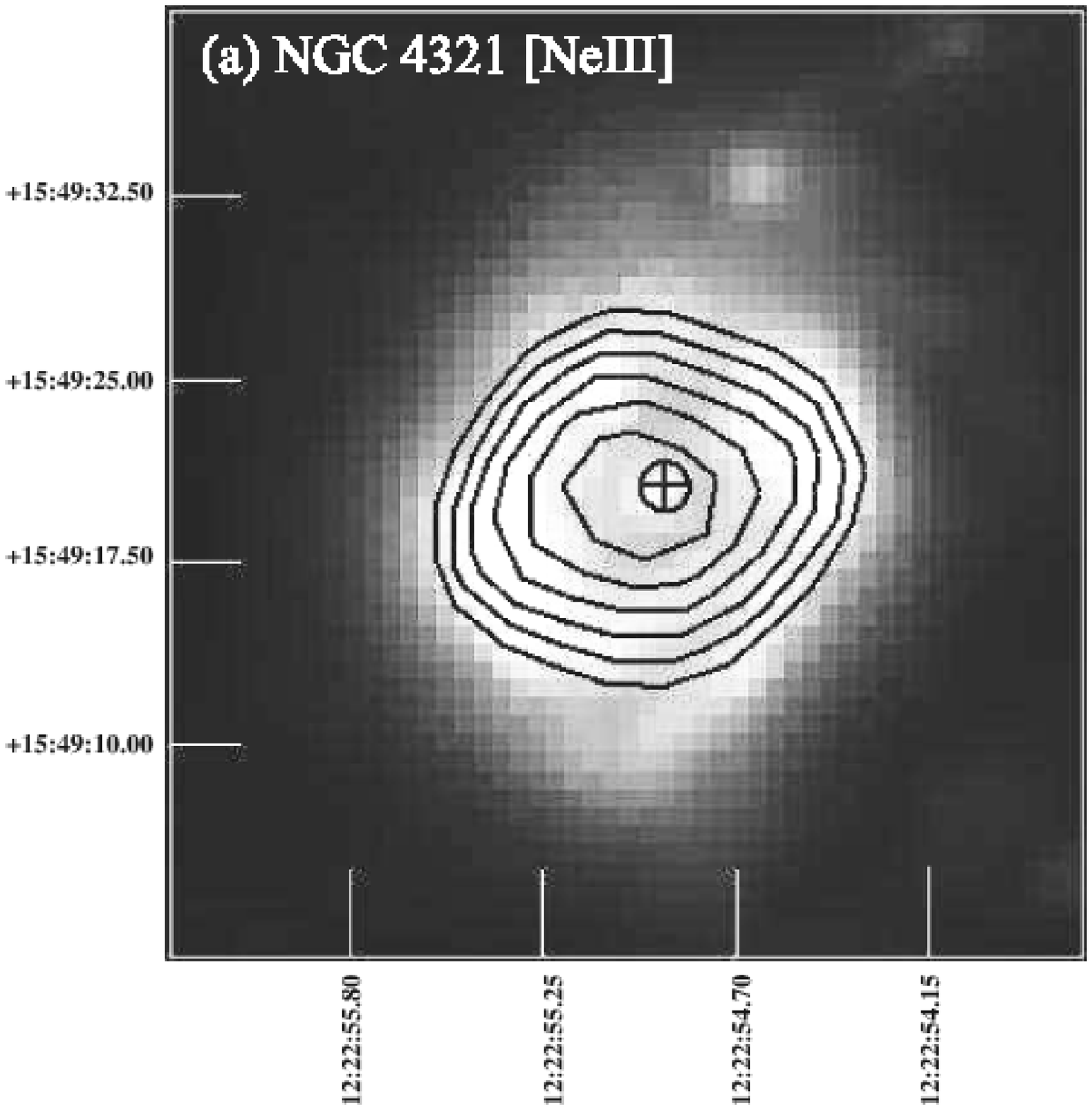} &
  \includegraphics[width=0.38\textwidth,angle=0]{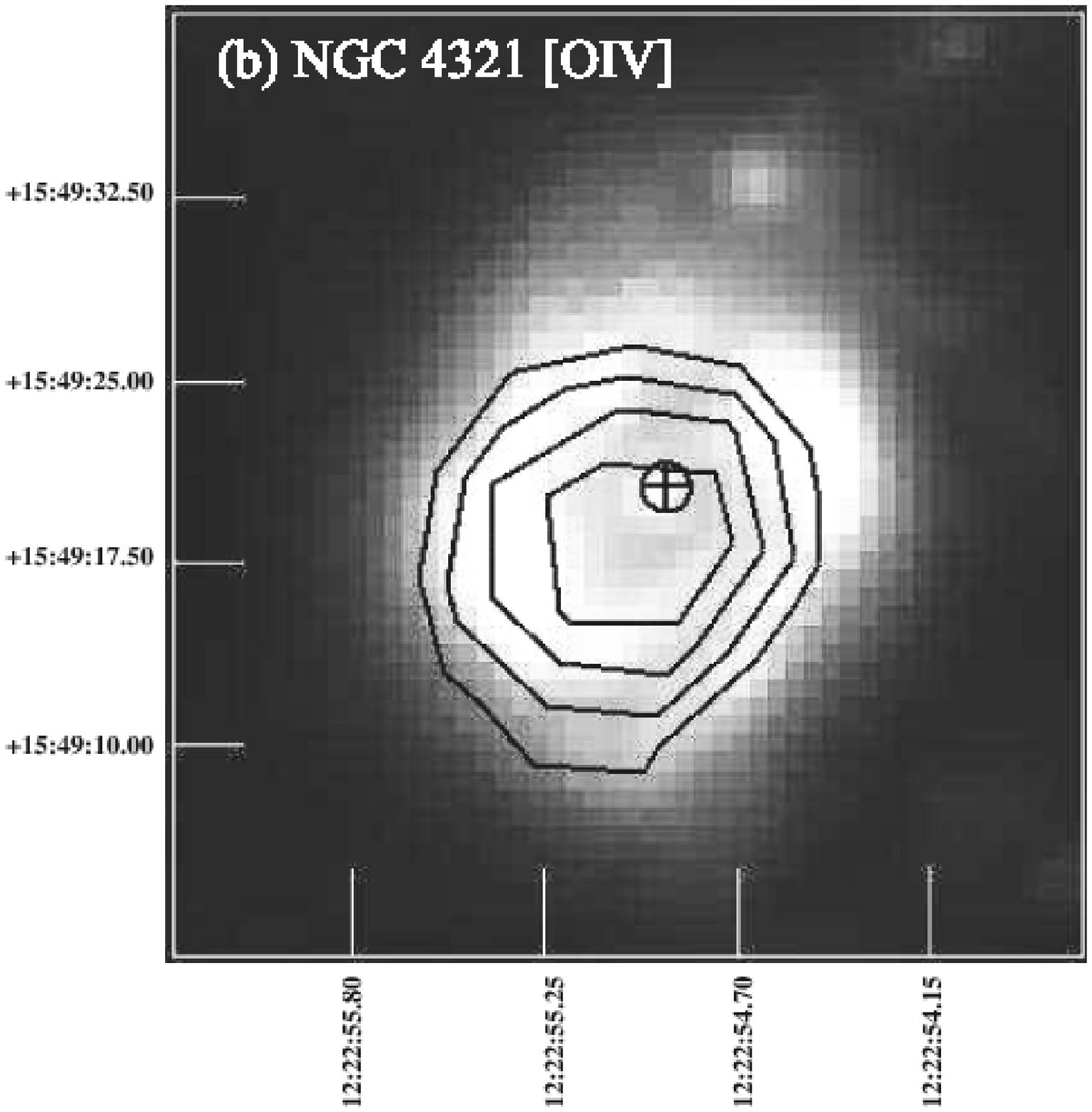} \\
  \includegraphics[height=0.38\textwidth,angle=0]{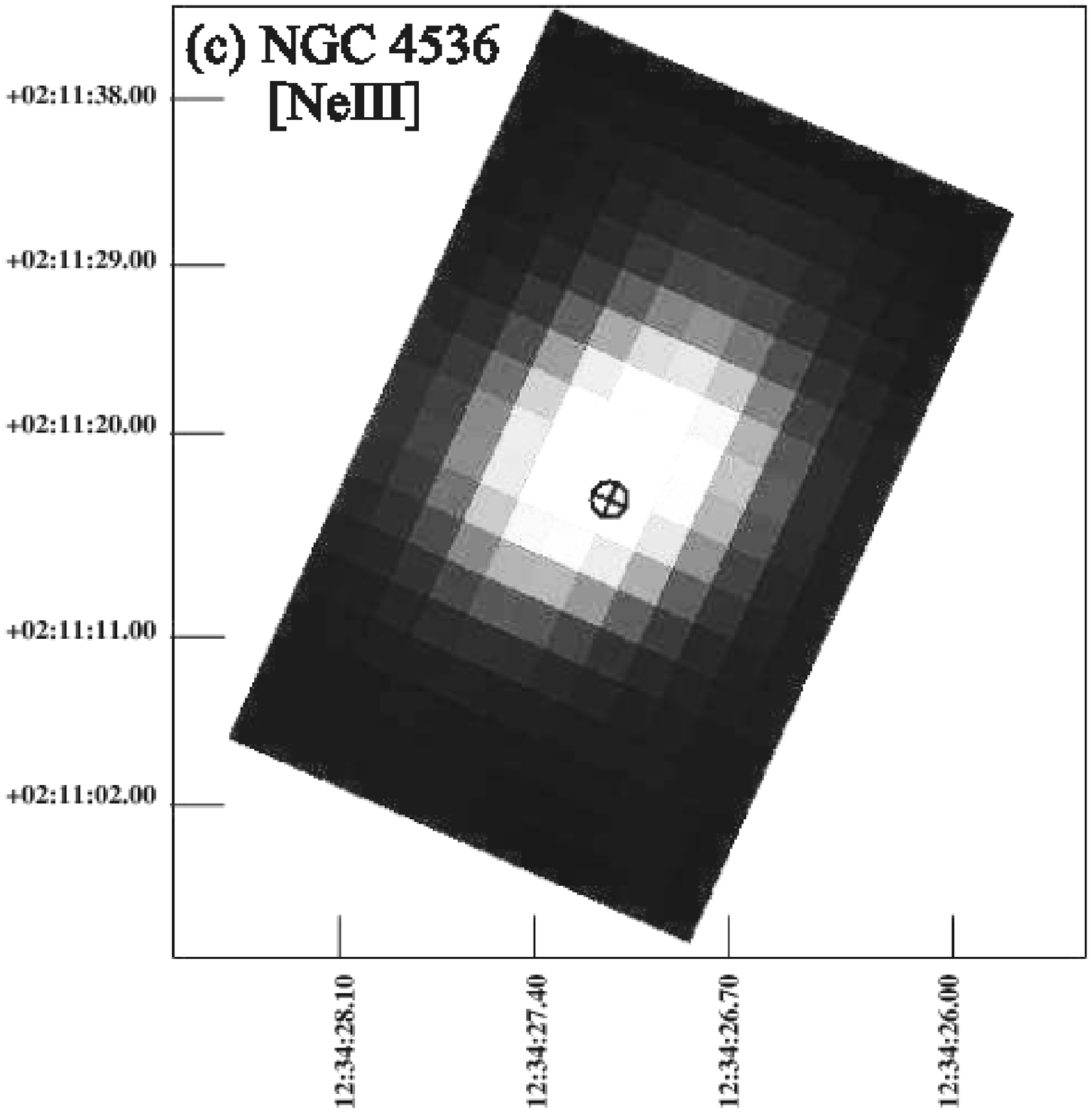} &
  \includegraphics[height=0.38\textwidth,angle=0]{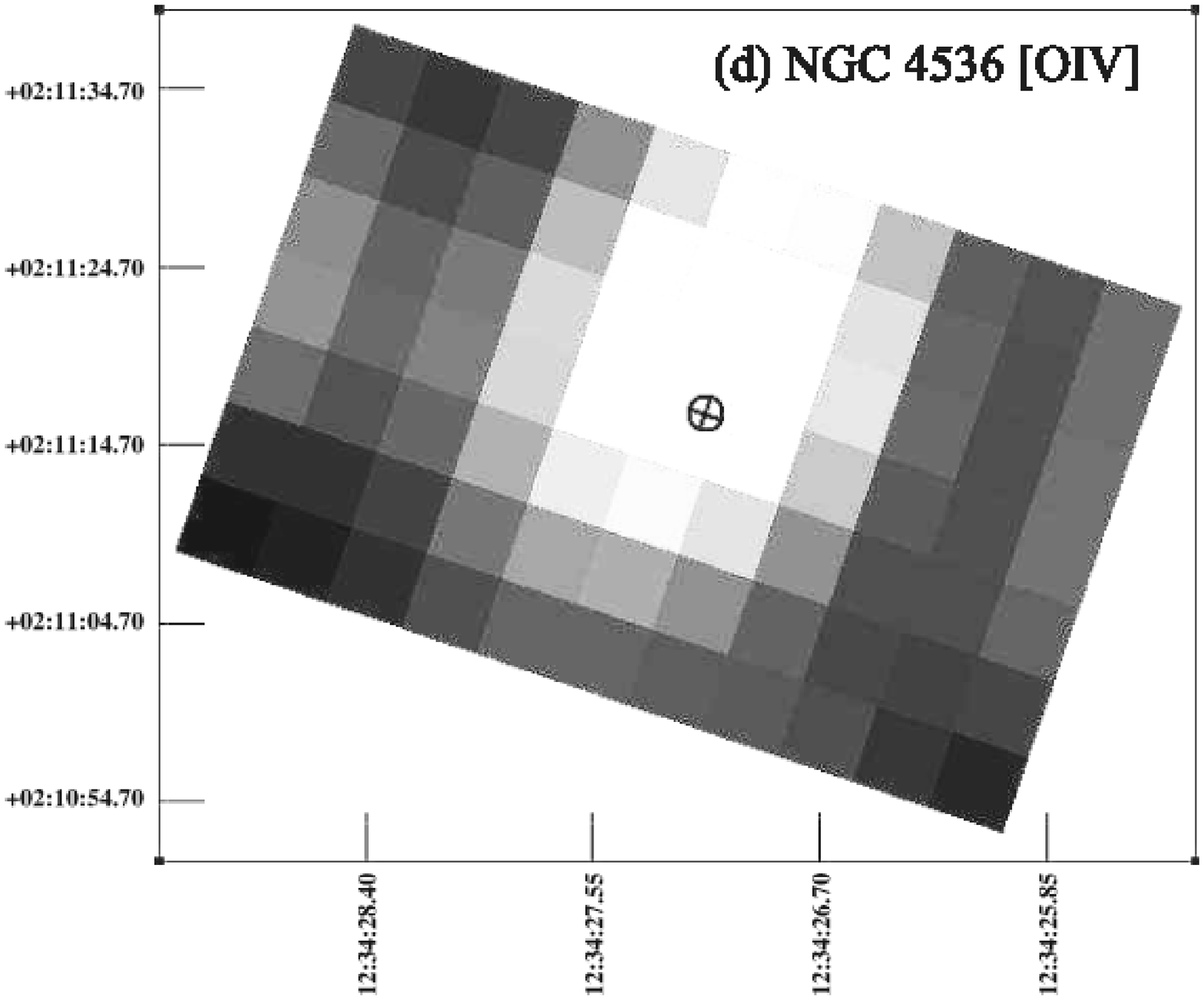} \\
  \includegraphics[height=0.38\textwidth,angle=0]{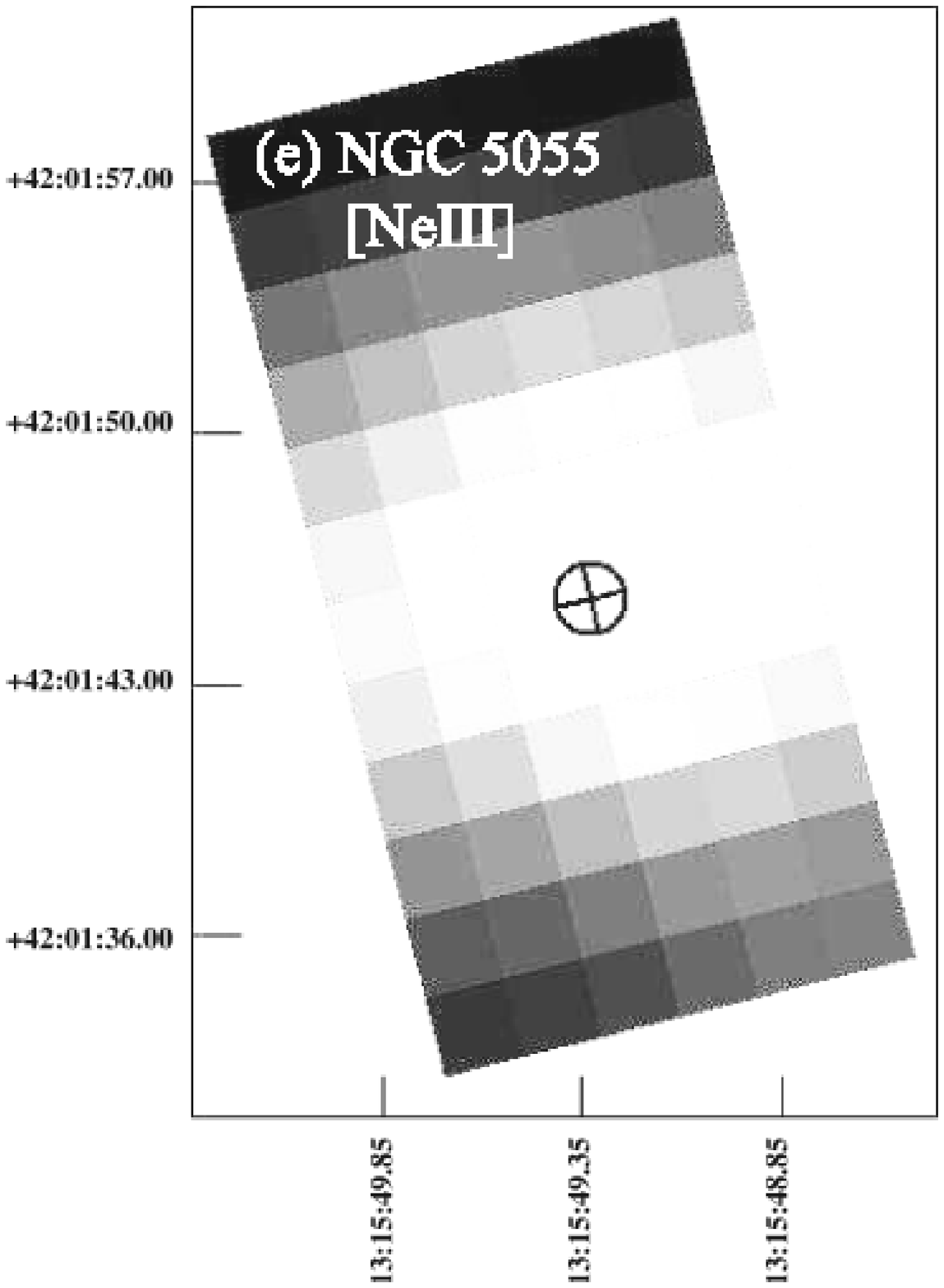} &
  \includegraphics[height=0.38\textwidth,angle=0]{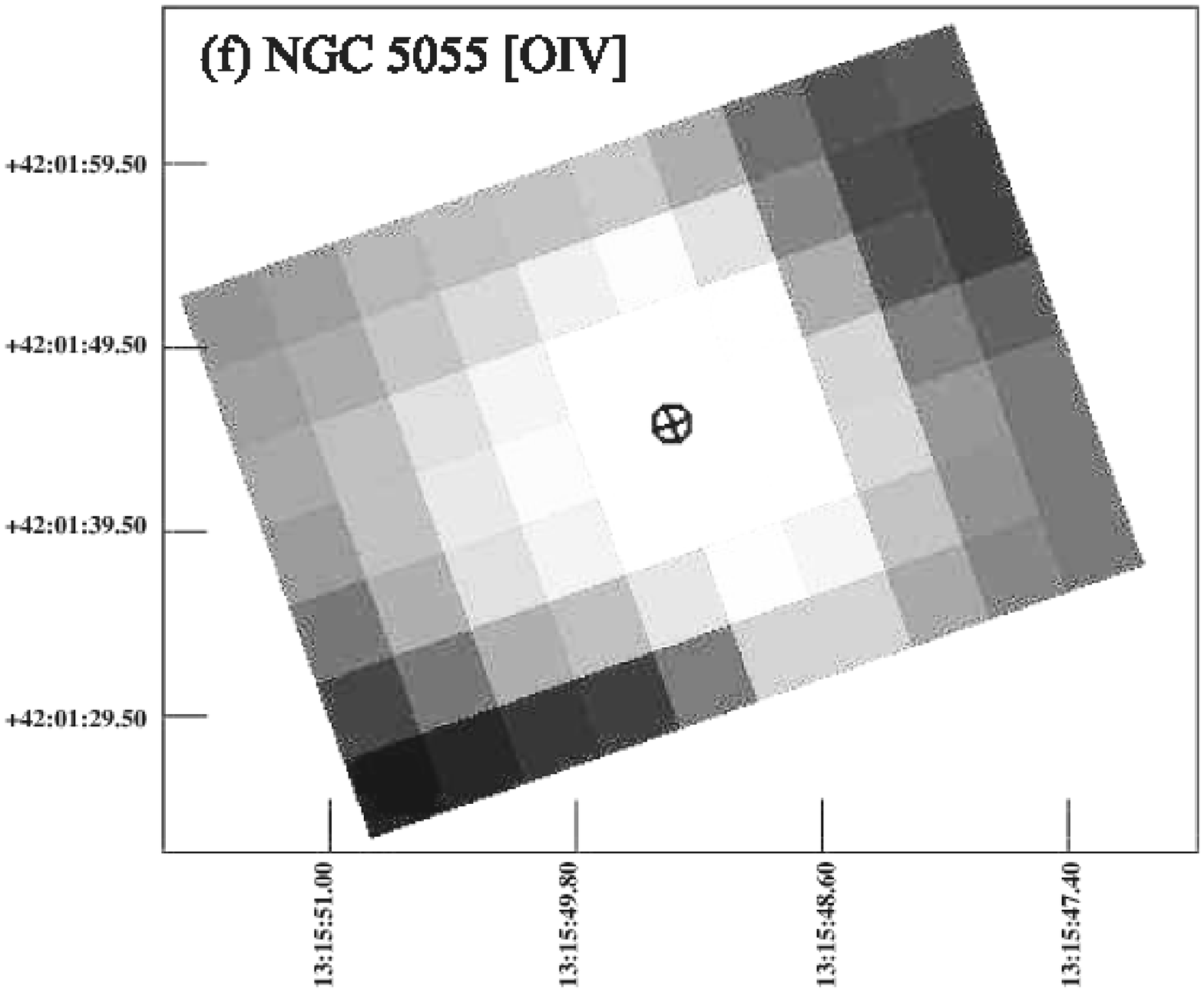} \\
\end{tabular}
\end{center}

\caption[]{(a) The [NeIII] 15.5$\mu$m continuum-subtracted image of NGC 4321. Contour overlays of the archival IRAC 8$\mu$m image are also shown and display the prominent circumnuclear star formation ring in this galaxy. Note that there are two components to the [NeIII] emission--one that follows the circumnuclear ring and a compact centrally concentrated component likely originated from gas ionized by the AGN.  (b) The [OIV] 26$\mu$m continuum-subtracted image of NGC 4321 with contour overlays of the [NeII] 12.8$\mu$m emission.  Note that the [NeII] emission traces the circumnuclear ring but the [OIV] emission is compact and centrally concentrated highly suggestive of an AGN origin.  (c) The [NeIII] 15.5$\mu$m continuum-subtracted image of NGC 4536. (d) The [OIV] 26$\mu$m continuum-subtracted image of NGC 4536. The emission appears to be associated with the nucleus however the [NeV] emission seems to be offset from the nuclear coordinates.  See Section 5.1 for details.  (e) The [NeIII] 15.5$\mu$m continuum-subtracted image of NGC 5055 (f) The [OIV] 26$\mu$m continuum-subtracted image of NGC 5055. The cross in all images indicates the nuclear coordinates from the 2MASS database. Note that in all images the emission is concentrated and centered on the nucleus, suggesting an AGN as the ionizing source.  }
\end{figure*}
%\clearpage
%\begin{center}
%\begin{tabular}{ccc}
 % \includegraphics[height=0.38\textwidth,angle=0]{f4c.ps} &
  %\includegraphics[height=0.38\textwidth,angle=0]{f4d.ps} \\
  %\includegraphics[height=0.38\textwidth,angle=0]{f4e.ps} &
  %\includegraphics[height=0.38\textwidth,angle=0]{f4f.ps} \\
  %\multicolumn{2}{c}{{\sc Fig} 4.--{\it Continued}}
%\end{tabular}
%\end{center}
%**********************************************************************

\section{ AGN Origin of the [NeV] Emission }

The detection of [NeV] emission from the 7 galaxies listed in Tables 3 and 4 is highly suggestive of the presence of an AGN in these galaxies. In this section, we summarize some of the published literature on each source highlighting any previous evidence for nuclear activity.  We also provide theoretical confirmation that the origin of the [NeV] emission is indeed an AGN.  Finally we determine the contribution of the AGN to the total luminosity.

\subsection{ Notes on Individual Galaxies }

{\bf NGC 3367:} NGC 3367 is an isolated face-on Sc barred galaxy (de Vaucouleurs et al. 1976) that is optically classified as an HII object.  Its optical line ratios place it well to the left of the Kewley et al. (2001) starburst theoretical limit line as can be seen in Figure 1, indicating that there is no hint of the presence of an AGN from its optical spectrum.  The aperture from which the [NeV] 14.3$\mu$m line was detected in this galaxy corresponds to a projected size of $\sim$ 1kpc $\times$ 2.4kpc. 

Although there is no hint of an AGN from optical observations, radio observations reveal a bipolar synchrotron outflow from a compact unresolved nucleus (of diameter less than 65 pc) and possibly two large lobes straddling the nucleus that extend up to $\sim$ 12kpc. (Garcia-Barreto et al. 1998).  There are no {\it Chandra} or {\it XMM} observations of NGC 3367 but it was detected by Einstein (Fabbiano et al. 1992).  The detection of the [NeV] lines in this work firmly establishes the existence of a weak AGN in this galaxy.

There are no published observations of the central stellar velocity dispersion in this galaxy.

{\bf NGC 3556:} NGC 3556 is an isolated edge-on Scd galaxy (de Vaucouleurs et al. 1976) that is also optically classified as an HII galaxy.  As can be seen in Figure 1, its optical line ratios are at the extreme low-end occupied by HII galaxies, well to the left of the starburst theoretical limit line from Kewley et al. (2001).  There is therefore absolutely no hint of the presence of an AGN in this galaxy in its optical spectrum. Adopting the distance to this galaxy from H97, the aperture from which the [NeV] 14.3$\mu$m line was detected corresponds to a projected size of $\sim$320pc$\times$770pc.

There is some previously published data that may be consistent with a weak AGN as well as strong nuclear star formation in this galaxy.  {\it Chandra} observations reveal numerous X-ray point sources scattered around the central regions with one source coincident with the optical nucleus that has a power-law X-ray spectrum typical for an AGN (Wang, Chaves, \& Irwin 2003). Prominent extraplanar diffuse X-ray emission is seen with substructure similar to that seen in H$\alpha$ images (Collins et al. 2000) possibly representing superbubbles of hot gas heated in massive star-forming regions.

There are no published observations of the central stellar velocity dispersion in this galaxy.

%**********************************************************************

\begin{figure}[]
\noindent{\includegraphics[width=9cm]{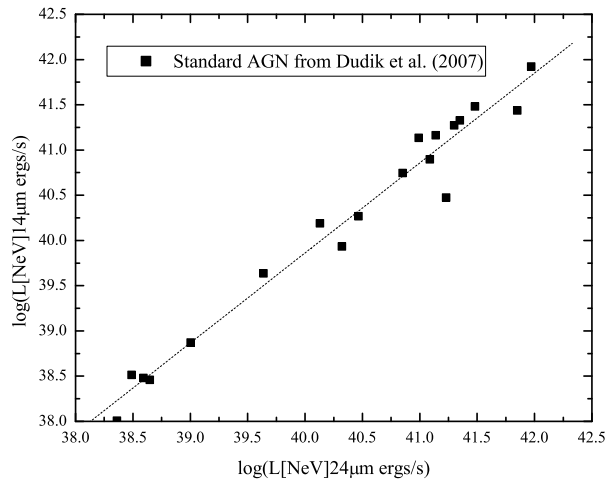}}
\caption[]{The [NeV] 14.3$\mu$m line luminosity as a function of the [NeV] 24.3$\mu$m line luminosity in known AGN that currently have [NeV] observations (Dudik et al. 2007).  The figure clearly shows a strong correlation. The Spearman rank correlation coefficient between the 14.3$\mu$m and 24.3$\mu$m line luminosities is $_{\rm S}$ = 0.97, with a probability of chance correlation of 6.7 $\times$ 10$^{-12}$, indicating a significant correlation.  The rms scatter in the relation is 0.219 dex. 
}
\end{figure}
%**********************************************************************

{\bf NGC 3938:} NGC 3938 is a nearly face-on Sc galaxy (de Vaucouleurs et al. 1976) at a distance of 17 Mpc (Tully \& Shaya 1984) that is also optically classified as an HII galaxy.  The [NeV] 24.3$\mu$m and the [OIV] 26$\mu$m lines were detected in this source in a $\sim$11\arcsec$\times$22\arcsec\, aperture which corresponds to a projected size of $\sim$ 0.9kpc$\times$1.8kpc.  

There are scattered HII regions in the central regions of this galaxy (Jiménez-Vicente et al. 1999) but no published evidence for AGN activity.  The nucleus was undetected at 6 cm (Ulvestad \& Ho 2002) and there are no published X-ray observations of this source.

The central stellar velocity dispersion in this galaxy is 40 km/s (Bottema 1988).

{\bf NGC 4321:}NGC 4321 is an Sbc spiral galaxy (de Vaucouleurs et al. 1976) at a distance of 16.8 Mpc (Tully \& Shaya 1984).  At this distance, the aperture from which the [NeV] 14.3$\mu$m line was detected corresponds to a projected size of $\sim$ 0.9kpc$\times$1.4kpc.  This galaxy is optically classified as a transition object with no broad lines detected (H97), implying that there is no firm evidence for an AGN based solely only its optical spectrum.  

Previously published multiwavelength observations reveal no signs of AGN activity in this galaxy .  There are several published radio observations displaying extended emission (Filho, Barthel \& Ho 2000, 2006), but no evidence for a compact flat spectrum radio core (Nagar et al. 2002).  The nucleus is not detected by {\it Chandra} and the upper limit to the 2-10 keV luminosity is L$_{\rm X}$ $\sim$ 3.6$\times$10$^{38}$ ergs 
s$^{-1}$ (Dudik et al. 2005) implying that the AGN is either weak or highly absorbed in the X-rays.  Mid-infrared (Wozniak et al. 1998), optical (Pierce 1986), H$\alpha$ (Knappen et al. 1995), and CO (Rand 1995) observations display a prominent circumnuclear ring of star formation activity which is also seen in the IRAC and [NeII] images (see Figure 4).  However, the emission from the higher ionization potential [NeIII] and [OIV] lines are clearly centrally concentrated and compact, strongly suggestive of AGN activity in the nucleus.  

The central stellar velocity dispersion in this galaxy is 83$\pm$12 km/s (Whitmore \& Kirshner 1981).

{\bf NGC 4414:} NGC 4414 is a relatively isolated flocculent Sc galaxy (de Vaucouleurs et al. 1976) that is optically classified as a transition object with no broad lines detected (H97).  There is therefore no firm evidence for an AGN based solely only its optical spectrum.  
  It's optical line ratios are very close to the Kewley et al.(2001) starburst theoretical limit line as can be seen in Figure 1.  

There are no published observations of NGC 4414 suggesting that it contains an AGN.  This source was not detected at radio frequencies by Filho, Barthel, \& Ho (2006) and arcsecond resolution VLA observations reveal a diffuse radio morphology (Filho, Barthel, \& Ho 2002).  There is also no indication of intense nuclear star formation.  Based on its infrared and H$\alpha$ luminosities, it is not classified as a starburst.  In fact, H$\alpha$  images show a central hole, indicating that there is no nuclear star formation (Pogge 1989) in this galaxy.  Although there is no evidence for powerful star formation activity, HI and CO observations indicate that it has one of the highest disk neutral surface gas densities known (Braine, Combes, \& van Driel 1993).

The central stellar velocity dispersion in this galaxy is 128 $\pm$ 9 km/s (Barth, Ho, \& Sargent 2002).

%**********************************************************************
\begin{table*}
\fontsize{9pt}{9pt}\selectfont
\begin{center}
\begin{tabular}{lcccccc}
\multicolumn{7}{l}{{\bf Table 4: Line Ratios and Luminosities of AGN Candidate Galaxies}}\\ 
\hline

\multicolumn{7}{l}{}\\
\multicolumn{1}{c}{Galaxy} & Projected  & \multicolumn{1}{c}{Projected} & \underline{[NeV] 14.32\micron} & \underline{[NeV] 14.32\micron} & \underline{[OIV] 25.89\micron} & L[NeV]\\

\multicolumn{1}{c}{Name} & \multicolumn{1}{c}{Aperture} & Aperture & [NeII] 12.81\micron & [NeIII] 15.56\micron &[NeII] 12.81\micron & 14.32\micron\\

 &\multicolumn{1}{c}{Size SH (kpc)} & Size LH (kpc) & & & & \\
\multicolumn{1}{c}{(1)} & (2) & \multicolumn{1}{c}{(3)} & (4) & (5) & (6) & (7)\\ 
\multicolumn{7}{l}{}\\
\hline
NGC3367 & .99x2.4 & 2.3x4.7 & 0.010$\pm$0.003 & 0.117$\pm$0.035 & 0.007$\pm$0.002 & 39.415\\
NGC3556 & .32x.77 & .76x1.5 & 0.017$\pm$0.004 & 0.126$\pm$0.030 & $<$0.063 & 	37.950\\
NGC3938\tablenotemark{\dag} & .39x.93 & .91x1.8 & 0.457$\pm$0.149 & $>$0.053 & 0.314$\pm$0.113 & 38.426\\
NGC4321 & 2.2x3.5 & 2.5x4.0 & 0.016$\pm$0.005 & 0.134$\pm$0.041 & $<$0.035 & 	38.913\\
NGC4414\tablenotemark{\dag} & .22x.53 & .52x1.0 & 0.154$\pm$0.048 & 0.391$\pm$0.123 & 0.361$\pm$0.050 & 37.980\\
NGC4536\tablenotemark{\ddag} & .58x.58 & .58x.58 & 0.012$\pm$0.003 & 0.062$\pm$0.017 & $<$0.047 & 	37.803\\
NGC5055 & .16x.39 & .39x.78 & 0.108$\pm$0.035 & 0.206$\pm$0.068 & 0.279$\pm$0.041 & 37.516\\
\hline
\end{tabular}
\end{center}
\tablecomments{{\bf Columns Explanation:} 
Col(1):  Common Source Names; 
Col(2):  Dimensions of the extraction region from the SH observation;
Col(3):  Dimensions of the extraction region from the LH observation;
Col(4):  Ratio of flux of [NeV] 14.32$\mu$m to [NeII] 12.81$\mu$m; uncertainties reported in columns 4, 5, and 6 are based on calibration uncertainties on the line flux of 30\% for mapping observations and 15\% for staring.   
Col(5):  Ratio of flux of [NeV] 14.32$\mu$m  to [NeIII] 15.56$\mu$m . 
Col(6):  Ratio of flux of [OIV] 25.89$\mu$m  to [NeII] 12.81$\mu$m ; upper limits provided for non-detections based on 3-sigma upper limit for [OIV].  
Col(7):  Log of luminosity of [NeV] 14.32$\mu$m  in units of erg s$^{-1}$.}
\tablenotetext{\dag}{Flux of [NeV] 14.32$\mu$m  estimated using [NeV] 24.32$\mu$m flux, according to equation 1, see section 4.3.}
\tablenotetext{\ddag}{[NeV] 14.32$\mu$m  emitting region does not overlap with {\it 2MASS} nuclear coordinates, See Table 2 for exact coordinates.}
\end{table*}
%**********************************************************************

{\bf NGC 4536:} NGC 4536 is a barred late-type spiral  (SABbc; de Vaucouleurs et al. 1976) which based on HST imaging, shows no evidence of a classical bulge but instead has a surface brightness profile consistent with a pseudobulge that seems to exhibit spiral structure (Fisher 2006).  This galaxy is optically classified by H97 as an HII galaxy.  We detected the [NeV] 14.3$\mu$m line in this galaxy from a $\sim$ 580pc$\times$580pc region (adopting the distance to NGC 4536 from H97) approximately 10\arcsec north-east of the optical nucleus of the galaxy.  This is the only galaxy in our sample for which the [NeV] emission seems to originate from a location that does not coincide with the optical nucleus as listed in NED.  The [OIV] 26$\mu$m and [NeIII] 15.5 $\mu$m emission however is centrally concentrated with a peak that is offset from the location of the [NeV] emission and  is closer to the optical nucleus than  defined in NED as can be seen from Figure 4.  The [OIV] 26$\mu$m line was not detected in the same aperture from which the [NeV] 14.3$\mu$m line was detected; the [OIV] flux is therefore listed as an upper limit in Tables 3 and 4 but is displayed in Figure 4.  We note that the sensitivity of the current mapping observation of NGC 4536 is insufficient to conduct a full exploration of the spatial morphology of the [NeV] emission in this galaxy.  Deeper IRS mapping observations are critical to confirm the exact peak of the [NeV] emission in this galaxy.

Apart from the [NeV] detection reported in this work, to the best of our knowledge, there is only one previously published observation of NGC 4536 that hints at the possibility of nuclear activity in this galaxy.  The optical line ratios obtained using recent high spatial resolution STIS spectroscopy suggests the presence of a weak AGN (Hughes et al. 2005). There is abundant evidence for powerful nuclear star formation in NGC 4536.  Its high far-infrared luminosity, strong H$\alpha$  (Pogee 1989), Br$\gamma$  (Puxley, Hawarden, \& Mountain 1988), and 10.8$\mu$m (Telesco, Dressel, \& Wolstencroft 1993) emission all suggest vigorous star formation in the central   $\sim$ 20\arcsec$\times$30\arcsec region.  The radio emission shows a diffuse morphology with 3 separate peaks (Vila et al. 1990, Laine et al. 2006), possibly representing an annular ring of star formation surrounding the nucleus.  However, Laine et al. (2006) suggest based on HST data that the radio clumps are close to the nucleus in projection but none correspond to the nucleus itself. This morphology is similar to that seen at 10.8$\mu$m (Telesco, Dressel, \& Wolstencroft 1993) as well as in the 1-0 S(1) molecular hydrogen line (Davies, Sugai, \& Ward 1997). 
We note that the variable resolution of most published observations, coupled with the fact that this galaxy is nearly edge-on and has an intricate central complex of emission at various wavelengths makes it very difficult to confirm the spatial coincidence of the various sources.   

 The galaxy was detected by Einstein (Fabbiano, Kim, \& Trinchieri 1992).  {\it ROSAT} HRI data reveal 2 ultraluminous X-ray sources (ULXs), one of which may be coincident with the optical nucleus (Ji-Feng \& Bregman 2005).   It is not possible based on existing X-ray observations to determine if this source is consistent with an AGN.  The galaxy was not observed by {\it Chandra}.  

Our observations provide strong motivation to conduct an extensive high-spatial resolution investigation at infrared and x-ray wavelengths of NGC 4536 to determine the exact location of the peak of the [NeV] emission and to confirm its association with an AGN in this galaxy.

The central stellar velocity dispersion in this galaxy is 84 $\pm$ 1 km/s (Batcheldor et al. 2005).

%**********************************************************************
\begin{figure}[]
\begin{center}
\noindent{\includegraphics[width=9cm]{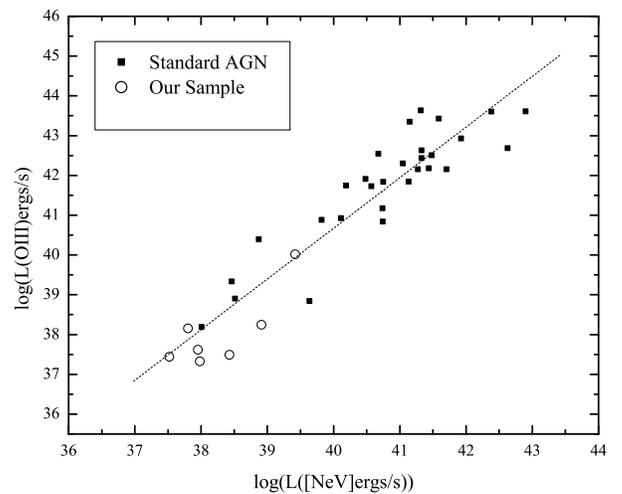}}
\end{center}
\caption[]{The [OIII] $\lambda$5007 line luminosity versus the [NeV]14.3$\mu$m  line luminosity for standard optically identified AGN and our sample of galaxies.  There is a significant correlation between the luminosities. The Spearman rank correlation coefficient between the line luminosities is $_{\rm S}$ = 0.84, with a probability of chance correlation of 8 $\times$ 10$^{-6}$, indicating a significant correlation.  The rms scatter in the relation is 0.67 dex.}
\end{figure}
%**********************************************************************
%**********************************************************************
\begin{figure}[]
\noindent{\includegraphics[width=9cm]{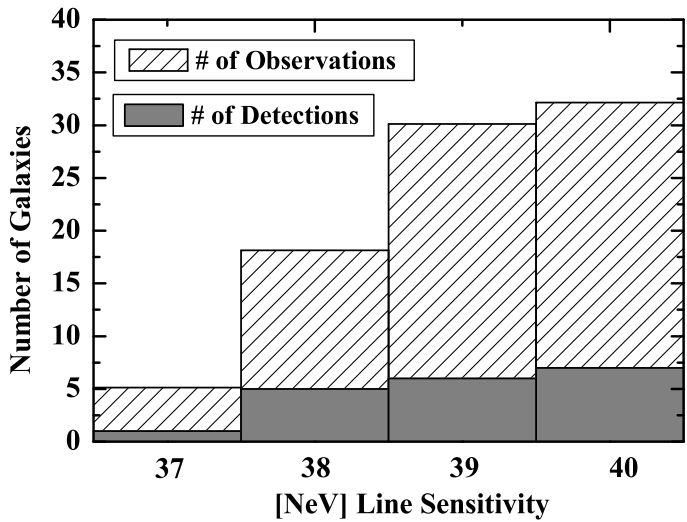}}
\caption[]{The distribution of galaxies with [NeV] detections as a function of limiting line sensitivity.  Plotted on the X-axis is the log of the 3$\sigma$ limiting [NeV] 14$\mu$m line sensitivity in units of ergs s$^{-1}$.  Galaxies are included in each bin if the sensitivity of the observations is equal to or better than the value listed.  Detections are indicated by the shaded histogram.  As can be seen, the detection rate at a limiting sensitivity of 10$^{38}$ ergs s$^{-1}$ is close to 30\%}
\end{figure}
%**********************************************************************

{\bf NGC 5055:} NGC 5055 is an Sbc spiral galaxy (de Vaucouleurs et al. 1976) at a distance of 7.2 Mpc (Pierce 1994).  At this distance, the aperture from which the [NeV] 14.3$\mu$m line was detected corresponds to a projected size of $\sim$170pc$\times$380pc.  This galaxy is optically classified as a transition object with no broad lines detected (H97).  Hence, there is no firm evidence for an AGN based solely only its optical spectrum.  

The evidence for AGN activity at other wavelengths is inconclusive.  The nucleus is  not detected at a number of radio frequencies except at 57.5 MHz with a flux density of 2.1 Jy (Israel \& Mahoney 1990), implying that the source is  radio-quiet.  A central UV source is detected, however high resolution {\it HST} observations show that it is resolved with a radius of $\sim$ 7 pc (Maoz et al. 1995) and shows no variability (Maoz et al. 2005).  The UV emission from this galaxy is therefore most likely dominated by a nuclear star cluster.  {\it Chandra} observations reveal the presence of numerous point sources scattered across the nuclear region and a bright hard point source coincident with the optical and infrared nucleus (Flohic et al. 2006; Luo et al. 2007).  Flohic et al. (2006) argue that the spectrum of this source is well-fitted by a two-temperature plasma model suggesting that the X-ray emission is powered by stellar processes.  Conversely, Luo et al. (2007) report that the central source is well-fitted by an absorbed power law with a spectral index typical of an AGN.  The [NeV] detection reported in this work provides firm confirmation of an AGN in this galaxy.

The central stellar velocity dispersion associated with the bulge in this galaxy is 103$\pm$6 km/s (Héraudeau.\& Simien 1998.)

\subsection{ Model Fits to the Mid-IR Fine Structure Line Fluxes: The Power of the AGN }

The mid-infrared fine structure line fluxes presented here can be modeled using photoionization models to confirm the AGN origin of the [NeV] emission and, in principle, estimate the contribution of the AGN to the bolometric luminosities in the 7 galaxies with [NeV] detections.  We use the spectral synthesis code {\it Cloudy} (Ferland et al. 1998) to model the mid-IR spectrum emitted by gas ionized by both an input AGN radiation field and a young starburst.  The modeling, along with a more extensive exploration of parameter space, is described in detail by Abel \& Satyapal (2007).  In short, we assume a typical AGN radiation field with a continuum shape characterized by the standard UV bump and X-ray power law as given in Korista et al. (1997).  The stellar continuum is chosen to be a 4 Myr continuous star-formation model with a Salpeter initial mass function (IMF) and star formation rate of 1 M$_{\odot}$${\rm yr^{-1}}$, generated using the {\it Starburst99}  website (Leitherer et al. 1999).  We have chosen conservatively to employ starburst parameters that generate the hardest ionizing radiation field.  This selection results in the maximum contribution of the starburst to the high-ionization line fluxes.  Our models assume a simple plane-parallel geometry with constant density typical of HII regions (log(n)=2.5 cm$^{-3}$).  The ionization parameter, U - the dimensionless ratio of ionizing flux to gas density - is varied, along with percent contribution of the AGN to the total luminosity.  The emission line ratios are most sensitive to changes in these parameters.  We emphasize that this is a simple model  , presented here with the intent of illustrating the effect the AGN luminosity has on the infrared and optical line ratios and to show that the mid-infrared line fluxes for our sample of 7 galaxies with [NeV] detections cannot be explained by pure photoionization by a starburst, even when an extreme starburst SED is adopted.

Figure 8 shows the predicted [NeV]14.3$\mu$m/[NeII]12.8$\mu$m flux ratio versus the [OI]/ H$\alpha$ and [SII]/ H$\alpha$ optical line flux ratios for varying values of U and AGN luminosity contribution, along with the observed values for all 7 galaxies.  We display only a narrow range of ionization parameters that generate line flux ratios within the range observed in our sample of galaxies.  A more extensive grid of theoretical calculations, with all standard optical line flux ratios plotted, is presented in Abel \& Satyapal (2007).  As expected, Figure 8 shows that when the ionization parameter is held constant, the [NeV]14.3$\mu$m/[NeII]12.8$\mu$m flux ratio increases dramatically as the AGN luminosity fraction increases.   Indeed, for logU=-2.5, the [NeV]14.3$\mu$m/[NeII]12.8$\mu$m flux ratio increases by over 5 orders of magnitude as the AGN luminosity fraction increases from 3\% to 100\%.  In contrast, the optical line flux ratios vary by only a factor of $\sim$ 10. Note that when the AGN luminosity fraction increases from 0.1\% to 10\%, {\it there is essentially no change in the optical line flux ratios}.  Figure 8 clearly shows that {\it infrared diagnostic ratios are much more sensitive to weak AGN than optical diagnostic ratios}.  This regime of parameter space is clearly best explored using the high ionization mid-infrared emission line fluxes presented in this work.

Surprisingly, when the AGN luminosity fraction goes to zero, [NeV] emission is still produced exclusively by the starburst.  The production of Ne4+ requires photons with energies $>$ 97eV, typically not produced by hot O-stars.  However, an extremely young starburst with a large population of Wolf-Rayet and O stars, can produce significant emission in the extreme ultraviolet (Schaerer \& Stasinka 1999; see Abel \& Satyapal (2007) for a more extensive discussion of this result ).  However, as can be seen from Figure 8, it is impossible to replicate the observed mid-infrared and optical line flux ratios in our sample of 7 galaxies with purely a starburst ionization radiation field.  An AGN contribution is required to explain the observed [NeV] emission in all cases.
%***********************************************
\begin{figure}[]
\begin{center}
  \includegraphics[width=0.5\textwidth]{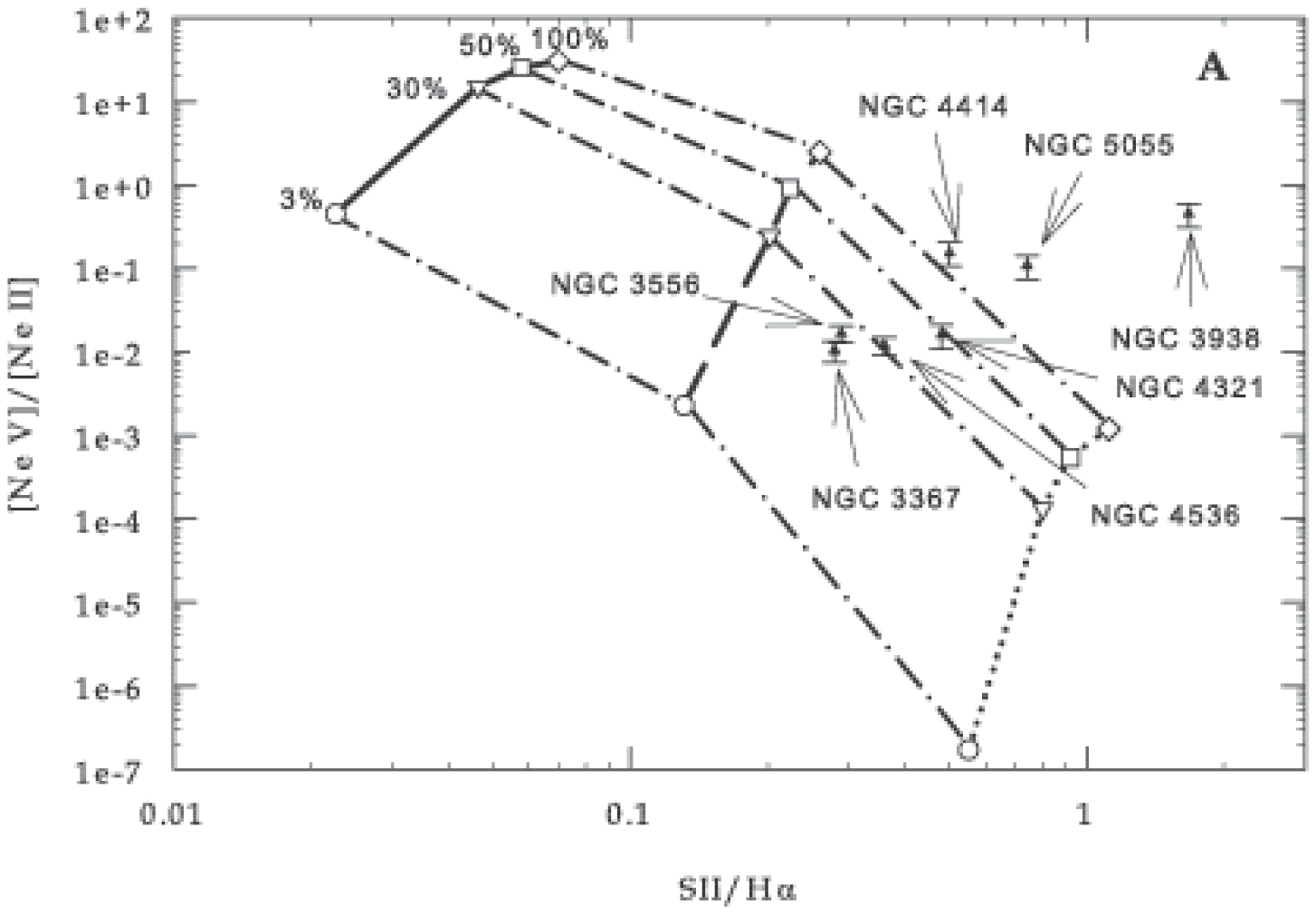}\\ 
  \includegraphics[width=0.5\textwidth]{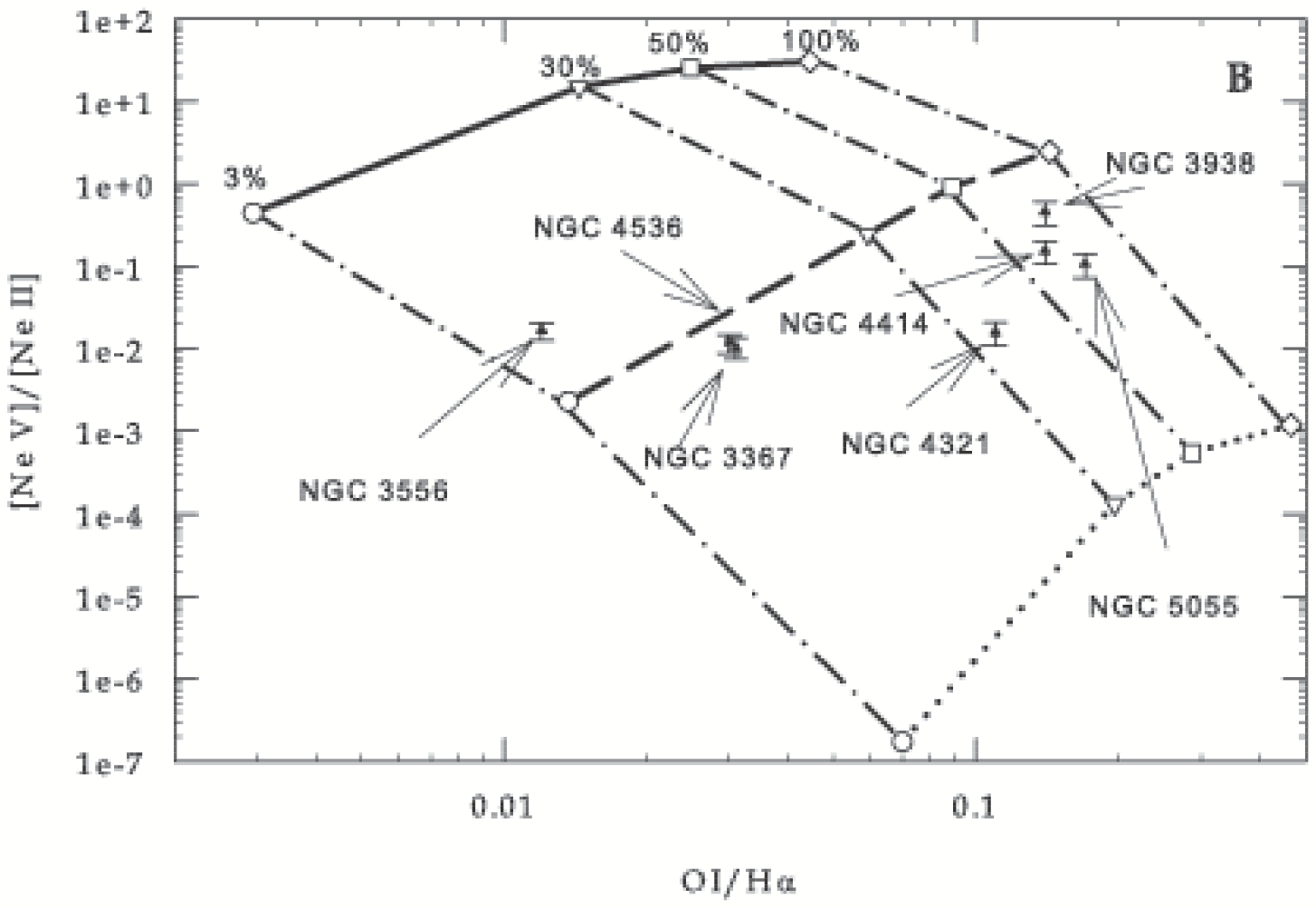}\\ 
\end{center}
\caption[]{The [NeV]14.3$\mu$m/[NeII]12.8$\mu$m line flux ratio versus the optical (a) [OI]/ H$\alpha$ flux ratio and (b) the [SII]/ H$\alpha$ flux ratio.  The solid, dashed, and dotted line display model results for ionization parameters of - 1.5, -2.5, and -3.5, respectively.  The dashed-dotted lines show the fraction of the total luminosity due to the AGN.  The line attached to the circles represent 3\% AGN, inverted triangles 10\% AGN, squares 30\% AGN, diamond 50\% AGN, and triangles 100\% AGN, as indicated in the Figure.  The 7 galaxies with [NeV] detections are also displayed.  We note that the optical line flux ratios are taken directly from Table 4 in H97.}
\end{figure}
%***********************************************

Figure 8 shows that 3 of the galaxies, NGC 3556, NGC 3367, and NGC 4536, appear to be dominated by star-formation, while the other four galaxies likely have a more substantial contribution from an AGN.  NGC 3556 could have as little as 10\% of its luminosity coming from the AGN.  The best-fit ionization parameter for our sample of galaxies lies between logU=- 2.5 and logU=-3.5.  The line flux ratios for NGC 4321 are consistent with a 30-50\% contribution from an AGN.  From Figure 8, it is clear that NGC 5055, NGC 3938, and NGC 4414 are not well-fit by our current model.  The line ratios in these objects cannot be replicated by changing the AGN SED, model geometry, or starburst SED, indicating that it is likely that in addition to photoionization by an AGN, shock-excitation of the narrow-line region (NLR) gas plays a significant role in these galaxies.  Indeed, an examination of several of the standard optically identified AGN with previously published mid-infrared spectroscopic observations (e.g. Sturm et al. 2002), shows that they too cannot be modeled solely using photoionization calculations, indicating that different physical mechanisms are at work in most AGN.  Clearly further  theoretical work, studying the combined effects of shocks and photoionization is needed.  In order to obtain a more precise estimate of the AGN fraction to the total luminosity, a model that includes the effects of shocks is required and will be explored in a future paper.  In this paper, we simply report that it is likely that NGC 4414, NGC 5055, and NGC 3938 likely have more dominant AGNs, but the actual fraction of the total luminosity due to the AGN in these galaxies in not well-determined using our current models.
  
\section{AGN Bolometric Luminosities and Black Hole Mass Estimates}

%***********************************************
\begin{table}
\fontsize{9pt}{9pt}\selectfont
\begin{center}
\begin{tabular}{lcccc}
\multicolumn{5}{l}{{\bf Table 5: Black Hole Properties}}\\ 
\hline

\multicolumn{5}{l}{}\\
\multicolumn{1}{c}{Galaxy} & log(L$_{BOL}$)  & log(M$_{EDD}$) & $\sigma$ & log(M$_{BH}$)\\

\multicolumn{1}{c}{Name} & \multicolumn{1}{c}{erg s$^{-1}$} & (M$_{\odot}$) & km s$^{-1}$ & (M$_{\odot}$)\\

\multicolumn{1}{c}{(1)} & (2) & \multicolumn{1}{c}{(3)} & (4) & (5)\\ 
\multicolumn{5}{l}{}\\
\hline
\multicolumn{5}{l}{}\\
NGC3367 & 43.29 & 5.19 & $\cdots$ & 7.84\\
NGC3556 & 41.91 & 3.81 & $\cdots$ & 7.38\\
NGC3938 & 42.36$^*$ & 4.26 & 40 & 4.82\\
NGC4321 & 42.82 & 4.72 & 83 & 6.36\\
NGC4414 & 41.94$^*$ & 3.84 & 128 & 7.28\\
NGC4536 & 41.78 & 3.68 & 84 & 6.39\\
NGC5055 & 41.51 & 3.41 & 103 & 6.82\\

\multicolumn{5}{l}{}\\
\hline
\end{tabular}
\end{center}
\tablecomments{{\bf Columns Explanation:}
Col(1):  Common Source Names; 
Col(2):  Bolometric luminosity in ergs/s estimated from the [NeV] 14 um luminosity using Equation (1) from Satyapal et al. (2007).  For L$_{BOL}$ marked with an asterisk (*), [NeV] 14 $\mu$m line luminosity estimated using the [NeV] 24 $\mu$m line luminosity;
Col(3):  Eddington mass luminosity calculated using estimated bolometric luminosity for the AGN;
Col(4):  Central stellar velocity dispersion (km/s).  References given for each galaxy in section 5.1;  
Col(5):  Estimate of black hole mass based on the stellar velocity dispersion when available (See Akritas \& Bershady 1996 and Ferrarese 2004) or the bulge magnitude for NGC 3556 and NGC 3367.}
\end{table}
%***********************************************

We can obtain order of magnitude estimates of the bolometric luminosity of the AGNs in the 7 galaxies with [NeV] detections using the [NeV] line luminosity. 
The starburst contribution to the [NeV] luminosity is negligible for our sample (Abel \& Satyapal 2007).  Assuming that the line emission arises exclusively from the AGN, we follow the procedure adopted by S07 to estimate the nuclear bolometric luminosity of the AGN.   Using the tight correlation between the [NeV] 14$\mu$m line luminosity and the AGN bolometric luminosity found in a large sample of standard AGN (Equation 1 in S07), we list in Table 5, the estimates of the AGN bolometric luminosity for the 7 galaxies with [NeV] detections.  This estimate assumes that the relationship between the [NeV] 14$\mu$m line luminosity and the bolometric luminosity established in more luminous AGN (see S07) extends to the lower [NeV] luminosity range for our sample.  The nuclear bolometric luminosities in our sample range from $\sim$ 3$\times$10$^{41}$ ergs s$^{-1}$ to $\sim$ 2$\times$10$^{43}$ ergs s$^{-1}$, with a median value of $\sim$ 9$\times$10$^{41}$ ergs s$^{-1}$.

If we make the assumption that the AGN is radiating at the Eddington limit, we can obtain a {\it lower mass limit} to the mass of the black hole.  In Table 5, we list the lower limits to the black hole mass based on the AGN bolometric luminosity estimates.  The Eddington mass estimates in our sample range from $\sim$ 3$\times$10$^3$M$_{\odot}$ to $\sim$ 1.5$\times$10$^5$M$_{\odot}$ with a median value of $\sim$ 7$\times$10$^3$M$_{\odot}$.

In order to determine if the lower mass limits derived for the black hole mass are incompatible with the M$_{BH}$-$\sigma$ relation, assuming a linear extrapolation to the mass range of our sample, we also list in Table 5, the expected black hole masses using the published central stellar velocity dispersions, when available, for our sources. In the absence of published central stellar velocity dispersion measurements, we list in Table 5 black hole masses expected using the bulge magnitude assuming the updated calibration of the Magorrian relationship from Ferrarese \& Ford (2005).  Estimates for the bulge magnitude in these cases are taken directly from Table 11 of H97 and are crude estimates based on the morphological type of the galaxy and its total luminosity.  In all cases, our lower mass limit is below the black hole mass estimate based on the M$_{BH}$-$\sigma$  relation, by a factor of $\sim$ 4 to three orders of magnitude, indicating that our results are not necessarily inconsistent with the M$_{BH}$-$\sigma$ relation.

\section{Summary and Conclusions}

We conducted a mid-infrared spectroscopic investigation of 32 late-type (Hubble type of Sbc or later) galaxies showing no definitive signatures of AGN in their optical spectra in order to search for low luminosity and/or embedded AGN.  The primary goal of our study was determine if AGN in low-bulge environments are more common than once thought.  Our high resolution {\it Spitzer} spectroscopic observations reveal that the answer to this question is {\it yes}.  Our main results are summarized below:

\begin{enumerate}

\item We detected the high ionization [NeV] 14.3$\mu$m and/or 24.3$\mu$m lines in 7 late-type galaxies, providing strong evidence for AGNs in these galaxies.  

\item We detected the high excitation [OIV] 25.9$\mu$m and [NeIII] 15.5$\mu$m lines in 5 out of the 7 of the galaxies with [NeV] emission.  Although these lines can be excited in star forming regions, our mapping observations (when available) suggest that the emission is centrally concentrated and likely to be dominated by the AGN.

\item Taking into account the range of  sensitivities of our observations, our work suggests that the AGN detection rate based on mid-infrared diagnostics in late-type optically normal galaxies can be as much or more than $\sim$ 30\%.  This detection rate implies that the overall fraction of late-type galaxies hosting AGN is possibly more than 4 times larger than what optical spectroscopic observations alone suggest.

\item Several of the galaxies with [NeV] detections have optical emission line ratios in the extreme ``starburst range'' indicating that there is absolutely no hint of an AGN based on their optical spectra. Three out of the 7 galaxies are classified based on their optical line ratios as ``transition objects'' but none show broad permitted optical lines.

\item Amongst the 7 AGN candidates in our sample, 3 are Sbc, 3 are Sc, and 1 is of Hubble type Scd.  Since there are only 3 galaxies of Hubble type Sd, our limited sample size precludes us from making any definitive conclusions on the incidence of AGN in completely bulgeless galaxies.  The lowest central stellar velocity dispersion amongst the galaxies with published measurements is 40 km/s.

\item We demonstrate using photoionization models with both an input AGN and an extreme EUV-bright starburst ionizing radiation field that the observed mid-infrared line ratios in our 7 AGN candidates cannot be replicated unless an AGN contribution is included. These models show that when the fraction of the total luminosity due to the AGN is low, the optical diagnostics are insensitive to the presence of the AGN.  In this regime of parameter space, the mid-infrared diagnostics offer a powerful tool in uncovering AGN missed by optical spectroscopy. 

\item Three of the galaxies, NGC 3556, NGC 3367, and NGC 4536, appear to be dominated by star-formation.  NGC 3556 could have as little as 10\% of its luminosity coming from the AGN. NGC 4321, NGC 5055, NGC 3938, and NGC 4414 likely have a more dominant contribution to their luminosity from the AGN in addition to having some contribution to their emission line fluxes from shock-excited gas.  All of the galaxies  that H97 classifies as HII galaxies are well characterized by a low AGN contribution, while all the transition objects seem to require some shock component. 

\item The AGN bolometric luminosities inferred using our [NeV] line luminosities range from $\sim$ 3$\times$10$^{41}$ ergs s$^{-1}$ to $\sim$ 2$\times$10$^{43}$ ergs s$^{-1}$, with a median value of $\sim$ 9$\times$10$^{41}$ ergs s$^{-1}$. Assuming that the AGN is radiating at the Eddington limit, this corresponds a lower mass limit for the black hole that ranges from  $\sim$ 3$\times$10$^3$M$_{\odot}$ to as high as $\sim$ 1.5$\times$10$^5$M$_{\odot}$. These lower mass limits however do not put a strain on the well-known relationship between the black hole mass and the host galaxy's stellar velocity dispersion established in predominantly early-type galaxies.  

\end{enumerate}

The {\it Spitzer} spectroscopic study presented here demonstrates that black holes do form and grow in low-bulge environments and that they are significantly more common than optical studies indicate.  In order to truly determine how common SBHs and AGN activity are in {\it completely} bulgeless galaxies, a more extensive study with {\it Spitzer} is crucial.

\acknowledgements
We are very thankful to Diana Marcu and Brian O'Halloran for their invaluable help in the data analysis and to Jackie Fischer, Mario Gliozzi, and Rita Sambruna for their enlightening and thoughtful comments.  We are also very grateful to Kartik Sheth and Daniel Dale for very helpful support with our questions about calibration and to the {\it Spitzer} helpdesk for numerous emails in support of our data analysis questions.  We are also very grateful for the helpful comments from the referee, which improved this paper.  This research has made use of the NASA/IPAC Extragalactic Database (NED) which is operated by the Jet Propulsion Laboratory, California Institute of Technology, under contract with the National Aeronautics and Space Administration.  SS gratefully acknowledges financial support from NASA grant NAG5-11432.  NA gratefully acknowledges NSF grant 0094050 and 0607497.  RPD gratefully acknowledges financial support from the NASA Graduate Student Research Program.

{}

\end{document}